\documentclass[amsmath,amssymb,12pt,superscriptaddress,nofootinbib]{revtex4}

\usepackage{graphicx}
\usepackage{dcolumn}
\usepackage{bm}
\usepackage{color}
\usepackage{slashbox}

\newcommand{\dd}{\mathrm{d}}

\newcommand{\del}{\partial}
\newcommand{\ee}{{\rm e}}

\newcommand{\ts}{t^{\rm s}}
\newcommand{\chis}{\chi^{\rm s}}
\newcommand{\rs}{r^{\rm s}}

\newcommand{\tb}{t_{\rm B}}

\definecolor{DarkBlue}{rgb}{0,0,0.7} 

\definecolor{DarkRed}{rgb}{0.65,0,0}

\begin{document}
\baselineskip5.5mm

\thispagestyle{empty}

{\baselineskip0pt
\leftline{
\baselineskip14pt
\sl\vbox to0pt{
              \hbox{\it DAMTP, Centre for Mathematical Sciences, University of Cambridge}
               \vspace{1mm}
                  \hbox{\it Division of Particle and Astrophysical Science, Nagoya University} 
                          \vss}}
\rightline{\baselineskip16pt\rm\vbox to11pt{
\vss}}%
}

\author{Norihiro~Tanahashi}\email{N.Tanahashi@damtp.cam.ac.uk}
\affiliation{
DAMTP, Centre for Mathematical Sciences, University of Cambridge, 
Wilberforce Road, Cambridge CB3 0WA, UK
}

\author{Chul-Moon~Yoo}\email{yoo@gravity.phys.nagoya-u.ac.jp}
\affiliation{
Gravity and Particle Cosmology Group,
Division of Particle and Astrophysical Science,
Graduate School of Science, Nagoya University, 
Nagoya 464-8602, Japan
}

\vskip1cm

\title{Spherical Domain Wall Collapse in a Dust Universe}

\begin{abstract}
\vskip0.5cm
To clarify observational consequence of bubble nucleations in inflationary era,
we analyse dynamics of a spherical domain wall in an expanding universe. %
We consider a spherical shell of the domain wall with tension $\sigma$ collapsing
in a spherically-symmetric dust universe, which is initially separated into the open Friedmann-Lema\^itre-Robertson-Walker universe inside the shell and the Einstein-de Sitter universe outside.
The domain wall shell collapses due to the tension, and sweeps the dust fluid. 
The
universe after the collapse becomes inhomogeneous and is 
described by the Lema\^itre-Tolman-Bondi model.
We construct solutions describing this inhomogeneous universe 
by solving dynamical equations obtained from Israel's junction conditions 
applied to this system.
We find that a black hole forms 
after the domain wall collapse
for any initial condition, and that the black hole mass at the moment of its formation is universally given by $M_{\rm BH}\simeq 17 \sigma/H_{\rm hc}$,
where $H_{\rm hc}$ is the Hubble parameter 
at the time when the shell radius becomes equal to the Hubble radius.
We also find that 
the dust fluid is distributed as $\rho\propto R^{3/2}$ near 
the central region after the collapse, where $R$ is the area radius.
These features would provide observable signatures of 
a spherical domain wall generated in the early universe.
\end{abstract}

\maketitle
\pagebreak

\section{Introduction}
The inflationary paradigm~\cite{Starobinsky:1980te,Sato:1980yn,Kazanas:1980tx,Guth:1980zm,Brout:1977ix} is one of crucial pieces in cosmology, and 
recent developments in observational cosmology
enabled us to study its features based on precision measurements.
For instance, 
the Planck Collaboration reported that 
some of phenomenological inflationary models 
are excluded at high confidence levels~\cite{Ade:2013uln}. 
However, there are still a lot of inflationary models that are consistent with current observations, and it is important to clarify observational features 
of those models to discriminate them.

Some inflationary models are based on plausible theories of high energy physics, such as the superstring theory and its low-energy effective theories.
The string landscape scenario is 
one of important concepts to discuss nature of the inflationary vacua 
in this class of models.
In this scenario,
quantum tunnelling is a key ingredient to create 
inflationary phase of our universe. 
Such a quantum tunnelling results in a bubble nucleation\cite{Coleman:1980aw} during inflation, and it may cause a multi-stream inflation~\cite{Li:2009sp,Duplessis:2012nb} whose signature would be observed in the CMB sky~\cite{Afshordi:2010wn}. 
The membrane nucleation in the context of the axion monodromy inflation~\cite{Kaloper:2008fb,Kaloper:2011jz} may provide another realization of the multi-stream inflationary model.

Let us consider a bubble nucleation with a spherical domain wall during inflation. 
Supposing that the regions inside and outside the domain wall are approximated by de Sitter solutions with different values of the cosmological constant, we can apply the membrane nucleation process proposed in Refs.~\cite{Brown:1988kg,Brown:1987dd}.
In this process,
the system is separated into two vacuum regions by a spherical domain wall at the moment of the bubble nucleation. 
The value of the cosmological constant corresponds to the potential height in each region, 
and the interior region has lower energy density 
compared to the exterior inflationary universe.
The domain wall becomes super-horizon size soon after the nucleation, and 
after the inflation end it comes back to our horizon and starts to collapse due to its tension.
In this paper, we analyze dynamics of such a domain wall shell just before and after it enters our horizon. 
Here, we assume that the domain wall shell survives without decaying even after the inflation, and that there is no interaction other than the gravitational one between the shell and matter fields during the collapse.

The collapsing domain wall will induce inhomogeneity to the background matter, and it would be an important observational feature of this system.
In general, the spacetime and matter after the shell collapse will be disturbed 
not only by the gravitational interactions but also by the sound wave 
emitted from the shell, 
and it would be difficult to study their dynamics without 
resorting to numerical simulations.
To circumvent this difficulty,
as a first step, we concentrate on a matter-dominated 
era and treat the matter component as dust fluid,
for which the sound speed vanishes.
Then, the spherically-symmetric inhomogeneous universe can be analytically described 
by the Lema\^itre-Tolman-Bondi (LTB) solution. 
This assumption puts a restriction on the applicable scope of our analysis. 
Since the bubble must enter the horizon during matter-dominated epoch, 
the domain wall size at the horizon crossing
must be larger than 100 Mpc comoving scale as long as we apply 
it to the late matter-dominated era.
Our analysis can be applied also to the matter-dominated era between the inflation end 
and the radiation-dominated era, if such an era exists, 
and the restriction on the domain wall size mentioned above 
could be avoided in this case.

Based on the discussions above, 
we study a spherical domain wall collapsing in a dust-dominated Friedmann-Lema\^itre-Robertson-Walker (FLRW) universe. The spacetime region swept by the domain wall becomes inhomogeneous and is described by the LTB solution. The region outside the initial comoving radius of the domain wall is not influenced by the domain wall collapse
and is described by the Einstein-de Sitter (EdS) universe.
Throughout this paper, 
we treat the domain wall as a thin shell with tension, that is, 
a shell with energy-momentum tensor equal to the induced metric 
multiplied by a constant coefficient. 
Applying Israel's junction conditions~\cite{Israel:1966rt} to the shell, we obtain a set of ordinary differential equations to fix the shell trajectory and the 
inhomogeneity in the LTB region.
We numerically solve these 
equations and 
figure out properties of the inhomogeneity.
We also find that a black hole forms at the center of the system in the final stage of the collapse, and study its properties.

This paper is organised as follows. 
After reviewing Israel's junction conditions briefly in section~\ref{junction}, 
we apply them to our system in section~\ref{regions}.
A set of ordinary differential equations 
for the shell trajectory and the LTB functions 
are derived in section~\ref{eqs}, and 
initial conditions for these equations are described in section~\ref{initial}. 
Results are listed in section~\ref{results}. 
Section~\ref{summary} is devoted to a summary. 

In this paper, we use the geometrised units 
in which the speed of light and Newton's gravitational constant 
are one.

\section{Junction Conditions with spherical symmetry}
\label{junction}
We consider a 3-dimensional spacelike or timelike hyper-surface $\Sigma$ 
embedded in a 4-dimensional spacetime $(\mathcal M, g_{\mu\nu})$. 
To fix our notation, we start with a brief review of Israel's junction conditions~\cite{Israel:1966rt} 
and their application to a system with spherical symmetry~\cite{Maeda:1985ye}.
For convenience, we first define the following brackets:
\begin{eqnarray}
\left[ A \right]^\pm&:=&A_+ - A_-, \\
\left\{ A \right\}^\pm&:=&A_+ + A_-, 
\end{eqnarray}
where 
we put subscripts $\pm$ to denote the value on each side of the hypersurface $\Sigma$. 
We also use the expression $\overline A$ defined by
\begin{equation}
\overline A:=\frac{1}{2}\left\{A \right\}^\pm=\frac{1}{2}\left(A_++A_-\right). 
\end{equation}

\subsection{Junction conditions}

Letting $s_\mu$ be unit normal form to $\Sigma$, 
we can define the induced metric $h_{\mu\nu}$ and the extrinsic curvature $K_{\mu\nu}$ 
as 
\begin{eqnarray}
h_{\mu\nu}&:=&g_{\mu\nu}-\epsilon s_\mu s_\nu, 
\\
K_{\mu\nu}&:=&h_\mu^{~\alpha}\nabla_\alpha s_\nu=D_\mu s_\nu, 
\end{eqnarray}
where $D_\mu$ is the covariant derivative on $\Sigma$ and 
$\epsilon=+1$ ($-1$) when $\Sigma$ is spacelike (timelike).
Then, the junction conditions are expressed as follows: 
\begin{itemize}
\item 
First junction condition
\begin{equation}
\left[ h_{\mu\nu}\right]^\pm=0. 
\label{eq1stj}
\end{equation}

\item
Second junction condition
\begin{equation}
\left[ K_{\mu\nu}\right]^\pm
=8\pi\epsilon\left(-S_{\mu\nu}+\frac{1}{2}S h_{\mu\nu}\right), 
\label{eq2ndj}
\end{equation}
where $S_{\mu\nu}$ is the energy momentum tensor of matter fields on $\Sigma$ and $S=h^{\mu\nu}S_{\mu\nu}$.
\item
Shell equation of motion 
\begin{equation}
S_{\mu\nu}\overline K^{\mu\nu}=\left[T_{\mu\nu}s^\mu s^\nu\right]^\pm, 
\label{seom}
\end{equation}
where $T_{\mu\nu}$ is the energy momentum tensor in 
the region $\mathcal M-\{\Sigma\}$. 
As is shown
in Ref.~\cite{Barrabes:1991ng}, the left-hand side of this equation
can be divided into (shell inertia) $\times$ (shell acceleration) and the surface pressure term,
while the right-hand side can be regarded as the net normal 
pressure on the surface from the outside. 
This equation is thus interpreted as the shell equation of motion.

\item
Shell energy conservation
\begin{equation}
D_\mu S^\mu_\nu =-\left[T_{\mu\alpha}s^\mu h^\alpha_\nu\right]^\pm. 
\label{seconv}
\end{equation}
This equation
can be derived by integrating 
the energy conservation law for the sum of $T_{\mu\nu}$ and $S_{\mu\nu}$
in the vicinity of the shell. 
The right-hand side corresponds to 
the energy-momentum flux into the shell from its surroundings.
\end{itemize}

\subsection{Spherically-symmetric spacetime and a shell}

In this subsection,
we apply the junction conditions 
summarised above to a spherically-symmetric spacetime and a shell. 
Hereafter, we assign a subscript $+$ ($-$) to 
variables in the region outside (inside) the shell. 
General spherically-symmetric line elements can be written as
\begin{equation}
\dd s^2=-\ee ^{2\alpha(t,\chi)}\dd t^2+\ee^{2\beta(t,\chi)}\dd \chi^2
+R^2(t,\chi)\left(\dd\theta^2+\sin^2\theta\dd \phi^2\right). 
\end{equation}
We consider the motion of a spherical shell in this spacetime
described by 
\begin{equation}
t=\ts(\tau),
\quad
\chi=\chis(\tau),
\label{shelltrajectory}
\end{equation}
where $\tau$ is the proper time associated with 
the shell trajectory in the radial direction.
The coordinate components of the radial tangent vector $v^\mu$ is given by 
\begin{equation}
v^\mu=(\dot \ts,~\dot \chis,~0,~0), 
\label{shelltangent}
\end{equation}
where the dot (\,$\dot~$\,) denotes a derivative with respect to $\tau$. 

We restrict the form of the energy momentum tensors 
$T_\pm^{\mu\nu}$ and $S^{\mu\nu}$ 
to the perfect fluid type, that is, 
\begin{eqnarray}
T^\pm_{\mu\nu}&=&\left(\rho_\pm+p_\pm\right)u^\pm_\mu u^\pm_\nu + p_\pm g^\pm_{\mu\nu},
\\
S^{\mu\nu}&=&(\sigma+\varpi)v^\mu v^\nu+\varpi h^{\mu\nu}, 
\end{eqnarray}
where $u_\pm^\mu$, $\rho_\pm$, $p_\pm$, $\sigma$ and $\varpi$
are the fluid four-velocity, energy density, pressure, 
shell surface density and surface pressure, respectively. 

Substituting these expressions into the junction conditions~(\ref{eq2ndj})--(\ref{seconv}), 
we obtain the following 4 nontrivial equations for the spherically-symmetric shell~\cite{Maeda:1985ye}:
\begin{itemize}
\item $(\theta,\theta)$ component of the second junction condition
\begin{equation}
\left[s^\mu \del_\mu \ln R\right]^\pm=-4\pi\sigma. 
\label{Kthetatheta}
\end{equation}

\item $(\tau,\tau)$ component of the second junction condition
\begin{equation}
\left[s_\mu \frac{Dv^\mu}{\dd \tau}\right]^\pm=4\pi (\sigma+2\varpi). 
\label{Ktautau}
\end{equation}

\item Shell equation of motion
\begin{equation}
\left\{s_\mu \frac{D v^\mu}{\dd \tau}\right\}^\pm
=-\frac{2}{\sigma}\left[(\rho+p)(u^\mu s_\mu)^2+p\right]^\pm
+\frac{2\varpi}{\sigma}\left\{s^\mu\del_\mu\ln R\right\}^\pm. 
\label{seom_sph}
\end{equation}

\item Shell energy conservation
\begin{equation}
D_\mu\left(v^\mu(\sigma+\varpi)\right)-v^\mu D_\mu \varpi=
\left[(p+\rho)u_\mu v^\mu u_\nu s^\nu\right]^\pm. 
\label{seconv_sph}
\end{equation}

\end{itemize}

\section{Shell in a dust-dominated universe}
\label{regions}

Based on the junction conditions for a spherically-symmetric spacetime, 
we consider dynamics of a domain wall shell in a dust-dominated FLRW universe
assuming spherical symmetry.
Due to the Ricci focusing effect of the shell, 
the expansion rate of the dust particle trajectories becomes smaller after the shell intersects them. 
This effect makes the density distribution of the dust fluid inhomogeneous in the shell exterior region,
and the resultant inhomogeneous universe is described by the LTB model.
We introduce geometric descriptions 
of this system
below.

\subsection{Shell interior: FLRW universe}
The shell interior region is filled by the FLRW universe with the metric given by
\begin{equation}
\dd s_-^2=-\dd t_-^2+a^2(t_-)\left[\dd \chi^2+f^2(\chi)\dd\Omega^2\right], 
\end{equation}
where 
\begin{equation}
f(\chi)=\left\{
\begin{array}{llll}
\sin\chi &{\rm for} &K=1& (\text{closed universe})\\
\chi &{\rm for} &K=0& (\text{flat universe})\\
\sinh\chi &{\rm for} &K=-1& (\text{open universe})
\end{array}
\right.. 
\end{equation}
Time evolution of the scale factor $a(t_-)$ is described by the Friedmann equations:
\begin{eqnarray}
8\pi\rho_-&=&3\left(\frac{\del_t a}{a}\right)^2+3\frac{K}{a^2},
\label{rhoeq_interior}
\\
0&=&-2\left(\frac{\del_t^2 a}{a}\right)-\left(\frac{\del_t a}{a}\right)^2
-\frac{K}{a^2}, 
\label{peq}
\end{eqnarray}
where we have omitted the subscript $-$ of $t_-$ in $\del_t a$ and $\del_t^2 a$ 
for notational simplicity. 
The energy-momentum tensor of the dust fluid is given by 
\begin{equation}
T_-^{\mu\nu}=\rho_-u_-^\mu u_-^\nu. 
\end{equation}

\subsection{Shell exterior: LTB model}

The spacetime outside the shell is described by the LTB model, whose metric is given by
\begin{equation}
\dd s_+^2=-\dd t_+^2+\frac{(\del_r R)^2}{1-k(r)r^2}\dd r^2
+R^2(t_+,r)\dd \Omega^2. 
\label{LTBmetric}
\end{equation}
where $k(r)$ is an arbitrary function. 
For this metric ansatz, the Einstein equations can be reduced to 
\begin{equation}
\left(\del_t R\right)^2=-kr^2+\frac{2M(r)}{R}, 
\label{Eeq}
\end{equation}
where $\del_t R:=\del_{t_+} R$.
$M(r)$ is the Misner-Sharp mass~\cite{Misner:1964je}
at the radius $r$, and it
is related to the dust energy density in this spacetime as 
\begin{equation}
8\pi\rho_+=\frac{2\del_r M}{R^2\del_r R}
=\frac{r^2m+\frac{1}{3}r^3\del_r m(r)}{R^2\del_r R}, 
\label{rhoeq_exterior}
\end{equation}
where we have defined $m(r)=6M(r)/r^3$. 
Hereafter, we describe the LTB solution using the expression given in Ref.~\cite{Tanimoto:2007dq}.
The solution for Eq.~\eqref{Eeq} is given by
\begin{equation}
R(t_+,r)=rm^{1/3}\bigl(t_+-t_{\rm B}(r)\bigr)^{2/3}S(x),
\end{equation}
where
\begin{equation}
x:=km^{-2/3}\left(t_+-t_{\rm B}\right)^{2/3}, 
\end{equation}
$\tb(r)$ is an arbitrary function that determines the big-bang time, 
and $S(x)$ is a function defined implicitly as
\begin{equation}
S(x)=
\left\{\begin{array}{lll}
\displaystyle
\frac{\cosh\sqrt{-\eta}-1}{6^{1/3}(\sinh\sqrt{-\eta}
-\sqrt{-\eta})^{2/3}}
\,;\qquad
&\displaystyle
x=\frac{-(\sinh\sqrt{-\eta}-\sqrt{-\eta})^{2/3}}{6^{2/3}}
\quad&\mbox{for}~~x<0\,,
\\
\displaystyle
\frac{1-\cos\sqrt{\eta}}{6^{1/3}(\sqrt{\eta}
-\sin\sqrt{\eta})^{2/3}}
\,;&\displaystyle
x=\frac{(\sqrt{\eta}-\sin\sqrt{\eta})^{2/3}}{6^{2/3}}
\quad&\mbox{for}~~x>0\,,
\end{array}\right.
\label{eq:defS}
\end{equation}
and $S(0)=({3}/{4})^{1/3}$. 
The function $S(x)$ is analytic for $x<(\pi/3)^{2/3}$, for which the metric~(\ref{LTBmetric}) describes a physical spacetime.
Some other characteristics of the function $S(x)$ are given in Refs.~\cite{Yoo:2008su,Tanimoto:2007dq}.
This expression of the LTB solution has an advantage for numerical analysis 
that the solution is described by apparently explicit functions of the coordinates 
$t,r$ once the value of $S(x)$ is determined numerically from Eq.~(\ref{eq:defS}).
The arbitrary functions $k(r)$, $m(r)$ and $t_B(r)$ will be determined according to the junction conditions and a gauge condition at the shell, as we see in the following sections.

\subsection{Shell trajectory}
A spherically-symmetric shell is separating the interior FLRW region and the exterior LTB region in our setup, and its trajectory is described by Eqs.~(\ref{shelltrajectory}) and (\ref{shelltangent}) in general spherically-symmetric coordinates.
We summarise their expressions in the coordinates introduced 
in the previous sections.

\begin{figure}[htbp]
\centering
\includegraphics[width=7cm,clip]{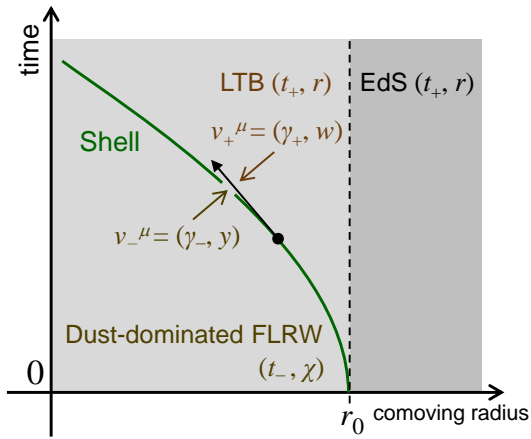}
\caption{
A schematic of our setup. 
The shell collapses in the dust-dominated FLRW universe, and the spacetime region after the shell collapse becomes inhomogeneous and is described by the LTB solution. The region outside the initial comoving radius of the shell $r_0$ is not influenced by the shell collapse and kept to be the EdS universe.
}
\label{Fig:geometry}
\end{figure}

The shell trajectory is described by the coordinates inside and outside the shell as
\begin{eqnarray}
t_-&=&\ts_-(\tau),\quad \chi=\chis(\tau), \\
t_+&=&\ts_+(\tau),\quad r=\rs(\tau), 
\end{eqnarray}
where $\tau$ is the proper time on the shell along the radial direction. 
The radial tangent vector $v_\pm^\mu$ in each coordinate is given by
\begin{eqnarray}
v_\pm^t&=&\dot t^{\rm s}_\pm,
\\
v_-^\chi&=&\dot \chi^{\rm s}=:y,
\label{defy}
\\
v_+^r&=&\dot r^{\rm s}=:w. 
\label{defw}
\end{eqnarray}
From the normalisation condition  $v_\pm^\mu v^\pm_\mu=-1$, we find 
\begin{eqnarray}
\dot t^{\rm s}_-
&=&
\sqrt{1+a^2y^2}
:=
\gamma_-
,
\label{defgammam}
\\
\dot t^{\rm s}_+&=&
\sqrt{1+\frac{\left(w\del_r R \right)^2}{1-kr^2}}
:=
\gamma_+.
\label{defgammap}
\end{eqnarray}
We show a schematic of the system and summarise the geometric quantities
in Fig.~\ref{Fig:geometry}.

We also introduce the unit radial vector $s_\pm^\mu$ normal to the shell, whose components are 
\begin{eqnarray}
(s^-_t,~s^-_\chi)&=&(-ay,~a \gamma_-),\\
(s_-^t,~s_-^\chi)&=&(a y,~\gamma_-/a), 
\end{eqnarray}
and 
\begin{eqnarray}
(s^+_t,~s^+_r)&=&\frac{\del_r R}{\sqrt{1-kr^2}}(-w,\gamma_+),\\
(s_+^t,~s_+^r)&=&\left(\frac{\del_r R}{\sqrt{1-kr^2}}w,
~\frac{\sqrt{1-kr^2}}{\del_r R}\gamma_+
\right). 
\end{eqnarray}
Using the normal vector $s_\pm^\mu$, 
the induced metric $h^\pm_{\mu\nu}$ is defined by 
\begin{equation}
h^\pm_{\mu\nu}=g^\pm_{\mu\nu}-s^\pm_\mu s^\pm_\nu. 
\end{equation}
Since the first junction condition requires $h^+_{\mu\nu}=h^-_{\mu\nu}$, 
we can regard $h_{\mu\nu}=h^\pm_{\mu\nu}$ as the unique 
metric on the shell. 

Motivated by the bubble nucleation scenario discussed in the introduction, 
we restrict the form of the shell energy momentum tensor $S_{\mu\nu}$ 
to the pure tension type, that is, 
\begin{equation}
S_{\mu\nu}=-\sigma h_{\mu\nu}
\end{equation}
with $\sigma$ being a constant.

\section{Evolution equations along the shell trajectory}
\label{eqs}

The shell trajectory and the arbitrary functions of the LTB model are determined by solving the junction conditions and the Einstein equations, which are summarised in sections~\ref{junction} and \ref{regions}.
We rewrite those basic equations into the form amenable to numerical integration.

Hereafter, we consider 9 ``dynamical" variables with 
respect to the time coordinate $\tau$:
$\ts_\pm(\tau)$, $\chis(\tau)$, $\rs(\tau)$, $y(\tau)$, $w(\tau)$, $m(\rs(\tau))$, $k(\rs(\tau))$ and $t_{\rm B}(\rs(\tau))$. 
We regard $m$, $k$ and $t_{\rm B}$ 
as functions of $\tau$ rather than $\rs$, and 
we omit the arguments of the functions and superscript ``s" 
for notational simplicity unless it is misleading.
We firstly fix the gauge freedom in $r$ by imposing the gauge condition~(\ref{rgauge}), and then describe the evolution equations obtained from the basic equations.

\begin{itemize}
\item 
Gauge condition on $r$
\begin{equation}
\del_r R=\sqrt{1-kr^2}.
\label{rgauge}
\end{equation}
We assume that this gauge condition is satisfied along the shell trajectory,
because this choice of gauge is useful to simplify the equations of motion derived below.
As shown in Ref.~\cite{Yoo:2010qn}, 
this condition can be rewritten as 
\begin{equation}
C\dot m+D\dot k +E\dot\tb=Fw
\label{deq1}
\end{equation}
with 
\begin{eqnarray}
C&=&\frac{1}{3}r m^{-2/3}(t-\tb)^{2/3}(S-2xS'), 
\label{eq:C}\\
D&=&r m^{-1/3}(t-\tb)^{4/3}S', 
\label{eq:D}\\
E&=&-\frac{2}{3}r m^{1/3}(t-\tb)^{-1/3}(S+xS'),
\label{eq:E}\\
F&=&\sqrt{1-kr^2}-m^{1/3}(t-\tb)^{2/3}S. 
\label{eq:F}
\end{eqnarray}
This equation is regarded as a dynamical equation 
since it contains derivative terms of dynamical quantities.  
The gauge freedom in $r$ is fixed by this dynamical equation 
once its initial value, which will be specified later, is given.

\item
First junction condition~(\ref{eq1stj})
\begin{equation}
af=R. 
\label{ceq1}
\end{equation}
Since this equation does not have any derivative of a dynamical quantity, 
it can be regarded as a constraint equation. 
Total differential of this equation with respect to $\tau$ 
gives another constraint equation as 
\begin{equation}
\gamma_- \del_t (\ln a)  +y \del_\chi (\ln f) 
=\gamma_+\del_t (\ln R)+w \del_r (\ln R). 
\label{ceq2}
\end{equation}

\item
Second junction condition
\begin{itemize}
\item $(\theta,\theta)$ component~(\ref{Kthetatheta})
\begin{equation}
w\del_t (\ln R)+\gamma_+\del_r (\ln R)
-y\del_t a-\frac{\gamma_-}{a}\del_\chi (\ln f)
=-4\pi\sigma.
\label{ceq3}
\end{equation}
This equation does not have any derivatives of dynamical variables, and thus it is 
 regarded as a constraint equation. 

\item $(\tau,\tau)$ component~(\ref{Ktautau})
\begin{equation}
\frac{\dot w}{\gamma_+}
-\frac{a\dot y}{\gamma_-}
+\frac{G\dot m}{\del_r R}+\frac{H\dot k}{\del_r R}
+\frac{I\dot \tb}{\del_r R}
+\frac{w J}{\del_r R}
-2\del_t a y
=-4\pi \sigma,
\label{deq4}
\end{equation}
\end{itemize}
where
\begin{eqnarray}
G&=&\del_t C=\frac{1}{6}r m^{-2/3}(t-\tb)^{-1/3}\frac{1}{S^2}, 
\label{eq:G}\\
H&=&\del_t D=\frac{2}{3}r m^{-1/3}(t-\tb)^{1/3}(2S'+xS''), 
\label{eq:H}\\
I&=&\del_t E=\frac{1}{6}r m^{1/3}(t-\tb)^{-4/3}\frac{1}{S^2}, 
\label{eq:I}\\
J&=&\frac{2}{3}m^{1/3}(t-\tb)^{-1/3}(S+xS'). 
\label{eq:J}
\end{eqnarray}

\item Shell equation of motion~(\ref{seom_sph})
\begin{eqnarray}
&&\frac{\dot w}{\gamma_+}
+\frac{a\dot y}{\gamma_-}
+\frac{G\dot m}{\del_r R}+\frac{H\dot k}{\del_r R}
+\frac{I\dot \tb}{\del_r R}
+\frac{w J}{\del_r R}
+2\del_t a y=-\frac{2}{\sigma}
\left(\rho_+w^2-\rho_-a^2y^2\right)\cr
&&\hspace{3cm}-2\left[w\del_t (\ln R)+\gamma_+\del_r (\ln R)+y\del_t a
+\frac{\gamma_-}{a}\del_\chi (\ln f)\right].
\label{deq5}
\end{eqnarray}

\item Shell energy conservation~(\ref{seconv_sph})
\begin{equation}
\rho_+\gamma_+w-\rho_-\gamma_-a y=0
\quad\Leftrightarrow \quad
\frac{\gamma_+r^2(3mw+r\dot m)}{24\pi R^2\del_r R}
-\rho_-\gamma_-a y=0. 
\label{deq6}
\end{equation}

\item Dynamical equations for $\ts_\pm$, $\chis$ and $\rs$

Equations  \eqref{defy}, \eqref{defw}, \eqref{defgammam}  and \eqref{defgammap}  
can be regarded as the dynamical equations for $\ts_\pm$, $\chis$ and $\rs$:
\begin{eqnarray}
\dot \ts_\pm&=&\gamma_\pm,
\label{deq7}
\\
\dot \chis&=&y, 
\label{deq8}
\\
\dot \rs&=&w.  
\label{deq9}
\end{eqnarray}
\end{itemize}

Up to this point,
we have found 8 dynamical 
equations (\eqref{deq1}, \eqref{deq4}, 
\eqref{deq5}, \eqref{deq6}, \eqref{deq7}, \eqref{deq8}, \eqref{deq9}) 
and 3 constraint equations (\eqref{ceq1}, \eqref{ceq2}, \eqref{ceq3})
in total.

From the constraint equations 
\eqref{ceq1}, \eqref{ceq2} and \eqref{ceq3}, 
we obtain the following expressions:
\begin{eqnarray}
R&=&af,
\label{exR}
\\
\del_t R&=&\del_t a f\gamma_-\gamma_++a\del_\chi f \gamma_+ y
-\del_\chi f \gamma_- w
-a\del_t a f y w
+4\pi \sigma af w ,
\label{exdtR}
\\
\del_r R&=&\del_\chi f \gamma_- \gamma_+ 
+a\del_t a f\gamma_+ y
-\del_ta f \gamma_-  w
-a \del_\chi f y w
-4\pi \sigma af\gamma_+. 
\label{exdrR}
\end{eqnarray}
Using Eqs.~\eqref{exdtR}, \eqref{exdrR} and the gauge condition~(\ref{rgauge}), 
the Einstein equation~(\ref{Eeq}) can be recast into the following form:
\begin{equation}
M=\frac{1}{6}mr^3=
\frac{4\pi}{3}a^3f^3\rho_-
+4\pi\sigma a^2 f^2 \left[\del_\chi f \gamma_- 
+a f \left(-2\pi \sigma +\del_t a y\right)\right]. 
\label{ceq4}
\end{equation}
The left-hand side is the quasi-local mass of the spacetime measured at the shell location on the outer side of it. In the right-hand side, this mass is decomposed into two terms, where the first one originates from the dust energy density $\rho_-$ contained inside the shell while the second one originates from the shell surface energy density $\sigma$. 
In the bubble nucleation scenario, the energy density inside the bubble decreases from that outside the bubble, and the energy deficit is converted into the shell surface energy. Equation (\ref{ceq4}) can be interpreted as an equation describing such a compensation of energy deficit by the shell surface energy.

Using Eqs.~\eqref{deq1}, \eqref{deq4}, \eqref{deq5}, \eqref{deq6}, we can obtain expressions for $\dot m$, $\dot k$, $\dot \tb$ and $\dot y$ in terms of $\dot w$ and quantities without derivatives. 
Using these expressions and the constraint equations, we can show that 
Eqs.~\eqref{ceq2} and \eqref{ceq3} are kept satisfied 
as long as they are satisfied at the initial time (see Appendix~\ref{App:constraints}).
Since only 8 dynamical equations are independent while there are 9 dynamical variables, we need another equation to determine their time evolution.
In this paper, we impose the following condition on 
the four-velocity of the dust fluid:
\begin{equation}
\left[ u^\mu s_\mu\right]^\pm=0 
\quad\Leftrightarrow \quad
w=ay. 
\label{ceq0}
\end{equation}
This condition prescribes that the dust particles pass through the shell without friction, that is, there is no interaction between dust particles and 
the shell except for the gravitational one.
This equation can be derived also from a requirement that the energy-momentum 
of the dust fluid and that of the shell are conserved individually.
Equation (\ref{seconv}) describes the conservation of the total energy-momentum, and by demanding both left and right-hand sides of this equation to vanish we can derive Eq.~(\ref{ceq0}).

Equation~\eqref{ceq0} can be regarded as an additional constraint equation. 
Combining this equation with Eq.~\eqref{deq6}, we find 
\begin{equation}
\rho_+=\rho_-. 
\label{equalrho}
\end{equation}
Also, differentiating Eq.~(\ref{ceq0}), we obtain an additional 
dynamical equation 
\begin{equation}
\dot w=\del_t a \gamma_- y+a\dot y. 
\label{deq0}
\end{equation}

In summary, we obtain
4 independent constraint equations and 9 independent dynamical equations
for 9 variables $t_\pm$, $\chi$, $r$, $y$, $w$, $k$, $m$, $\tb$. 
There are several trivial relations between these variables, 
and thus we do not need to solve all of the dynamical equations. 
For example, since 
\begin{equation}
\gamma_-=\gamma_+
\end{equation}
follows from Eq.~(\ref{ceq0}), we obtain 
\begin{equation}
\dot t_-=\dot t_+.
\label{tmtp}
\end{equation}

In the following, we rewrite the dynamical equations to 
the form convenient for numerical integration. 
First, we rewrite all the equations taking \eqref{ceq0} into account. 
Equations \eqref{exdtR} and \eqref{exdrR} can be rewritten as 
\begin{eqnarray}
\del_tR&=&f\left(\del_t a +4\pi a^2y\sigma\right), 
\label{exdtR-2}
\\
\del_rR&=&\del_\chi f -4\pi a f \gamma_- \sigma. 
\label{exdrR-2}
\end{eqnarray}
Equations \eqref{deq1}, 
\eqref{deq4}, \eqref{deq5} and \eqref{deq6} can be rewritten respectively as
\begin{eqnarray}
&&C\dot m+D\dot k +E\dot \tb =F a y, 
\label{deq1-2}
\\
&&
\frac{G\dot m}{\del_rR}
-
\frac{a\dot y }{\gamma_-}
+\frac{1}{\gamma_-}
\dot w 
+
\frac{H\dot k }{\del_rR}
+
\frac{I\dot \tb}{\del_rR}
+4\pi\sigma-2\del_t a y +\frac{ayJ}{\del_rR}=0, 
\label{deq4-2}
\\
&&
\frac{G\dot m}{\del_rR}
+
\frac{a\dot y }{\gamma_-}
+\frac{\dot w }{\gamma_-}
+
\frac{H\dot k }{\del_rR}
+
\frac{I\dot \tb}{\del_rR}
+\frac{4\del_\chi f \gamma_-}{af}-8\pi\sigma+6\del_t a y 
+\frac{ay J}{\del_rR}=0, 
\label{deq5-2}
\\
&&
r^3\dot m +3 a y m r^2 - 24 \pi a^3 f^2 y \del_rR \rho_- =0. 
\label{deq6-2}
\end{eqnarray}
From these equations and Eq.~\eqref{deq0}, we obtain
\begin{eqnarray}
\dot m &=&-\frac{3 a y m}{r} +\frac{24\pi a^3 f^2 y \del_rR \rho_-}{r^3}, 
\label{eq:dotm}
\\
\dot y &=&
-\frac{2\del_\chi f \gamma_-^2}{a^2 f}
+\frac{6\pi\gamma_- \sigma}{a}
-\frac{4\del_t a y \gamma_- }{a}, 
\label{eq:doty}
\\
\dot w &=&\del_t a \gamma_- y +a
\dot y
, 
\label{eq:dotw}
\\
\dot k &=&
\frac{8\pi\del_rR\del_tR\sigma}{r^2}
-\frac{2\del_tay\del_rR\del_tR}{r^2}
-\frac{2ayk}{r}
-\frac{y \del_rR m r}{3af^2}
+\frac{8\pi a^2 f y \del_rR \rho_-}{r^2}, 
\label{eq:dotk}
\\
\dot t_{\rm B}&=&
\frac{2D\del_rR}{r^2}\left(4\pi\sigma-\del_ta y\right)
+\frac{6ay}{r^3}\left(m-\frac{8\pi a^2f^2\del_rR\rho_-}{r^2}\right)\left(CH-DG\right)
\cr
&&+\frac{2ay}{r^2}\left(DJ+FH\right). 
\label{eq:dottb}
\end{eqnarray}
These equations and Eqs.~\eqref{deq7}, \eqref{deq8} and \eqref{deq9} 
constitute the dynamical equations for our system.
Since $w$ can be calculated from $y$ via Eq.~\eqref{ceq0}, 
we do not need to solve Eq.~\eqref{eq:dotw}. 
In addition, $t_+$ is trivially obtained from the value of $t_-$ and initial conditions
using Eq.~(\ref{tmtp}). 
We can also avoid solving Eqs.~\eqref{eq:dotk} and \eqref{eq:dottb} as follows.
The value of $k$ can be evaluated by Eq.~\eqref{Eeq} and the constraint equations. 
For $\tb$, we use integral of Eq.~\eqref{Eeq}
with respect to $R$ to express it in terms of $t_+$ and other dynamical variables:
\begin{equation}
t_+-\tb=
\left\{
\begin{array}{lll}
-\frac{\sqrt{R(mr-3kR)}}{\sqrt{3}kr}+\frac{m\arctan\left(\frac{\sqrt{3kR}}{\sqrt{mr-3kR}}\right)}{3k^{3/2}} & {\rm for}&\del_tR>0,  
\\
\frac{m\pi}{6k^{3/2}}+\frac{\sqrt{R(mr-3kR)}}{\sqrt{3}kr}
-\frac{m\arctan\left(\frac{\sqrt{3kR}}{\sqrt{mr-3kR}}\right)}{3k^{3/2}} & {\rm for}&\del_tR<0.  
\end{array}
\right.
\end{equation}

For our purpose, 
it is better to use $t_-$ or the scale factor $a$ 
as the variable to 
parametrise the shell trajectory instead of the proper time $\tau$, because 
$\tau$ degenerates if the trajectory approaches a 
null trajectory. 
$t_-$ is obtained from the integral of Eq.~\eqref{rhoeq_interior} with respect to $a$ as
\begin{equation}
\frac{t_-}{\alpha}=\sqrt{\frac{a}{\alpha}\left(1+\frac{a}{\alpha}\right)}
-\text{arcsinh} \sqrt{\frac{a}{\alpha}}, 
\label{}
\end{equation}
where  
\begin{equation}
\alpha=\frac{8}{3}\pi\rho_{-}a^3=\text{constant},
\end{equation}
and we have set $K=-1$, which holds in cases we are 
interested in as we see in section~\ref{initial}.
Below,
we use the scale factor $a$ as the independent variable based on these equations.

To summarise, we solve the following 4 differential equations in practice:  
\begin{eqnarray}
\frac{\dd \chi}{\dd a}&=&\hat y:=\frac{y}{\del_ta \gamma_-}, 
\\
\frac{\dd \hat y}{\dd a}&=&
-\frac{2\del_\chi f}{a^2(\del_ta)^2 f\gamma_-^2}
+\frac{6\pi \sigma}{a(\del_ta)^2\gamma_-^3}-\frac{4\hat y }{a\gamma_-^2}
-\frac{\del_t^2a\hat y}{(\del_ta)^2}-a(\del_ta)^2 \hat y^3, 
\\
\frac{\dd r}{\dd a}&=&\frac{w}{\del_ta\gamma_-}
=a\hat y
, 
\\
\frac{\dd m}{\dd a}&=&
-\frac{3a\hat y m}{r}+\frac{24\pi a^3f^2\hat y \del_rR \rho_-}{r^3},
\label{eq:dmda}
\end{eqnarray}
and determine the other variables using the equations derived above.

\section{Boundary conditions}
\label{initial}

Having clarified the evolution equations, 
we now need to supply the conditions on the initial time surface 
and at the outer boundary to determine evolution of the dynamical variables.

We assume that the shell comoving radius is not increasing at the initial time. 
Then, we can divide the spacetime into three domains: 
the dust FLRW region inside the shell, 
the LTB region outside the shell and 
the region outside the initial shell comoving radius. 
Since recent observations strongly suggest 
that our universe is very close to a spatially flat one, 
we assume the spacetime in this outermost region to be given by 
the dust-dominated flat (EdS) universe. 
This assumption will fix the boundary condition at the outer boundary
as discussed in section~\ref{Sec:BC} in more detail. 
We also describe the conditions on the initial time surface in section~\ref{Sec:IC}.

\subsection{Outer boundary}
\label{Sec:BC}

We summarise the boundary conditions on the outer boundary that separates 
the LTB region and the outermost EdS region in this section. 
We assume that there is no singular energy distribution on this boundary 
and that the boundary is comoving. 
From this requirement and the first junction condition, 
we find that the expansion rate must be continuous there:
\begin{equation}
\left(\frac{\del_tR(t_+,r_0)}{R(t_+,r_0)}\right)^2
=: H_+^2(t_+,r_0)
=H_{\rm EdS}(t_+)^2=
\frac{8\pi}{3}\rho_{\rm EdS}(t_+)
, 
\label{contH}
\end{equation}
where $r_0$, $H_{\rm EdS}$ and $\rho_{\rm EdS}$ are 
the comoving coordinate $r$ at the outer boundary,  
the Hubble expansion rate and the energy density of the EdS region, respectively. 
In addition, under this assumption, 
an equation similar to (\ref{ceq4}) implies
\begin{equation}
M(r_0)=\frac{4\pi}{3}R(t_+,r_0)^3\rho_{\rm EdS}(t_+),
\label{EdSmass}
\end{equation}
that is, the total mass surrounded by the outer boundary 
must be equal to that of the EdS spacetime with 
the same areal radius. 
From Eqs.~\eqref{Eeq}, \eqref{contH} and \eqref{EdSmass}, 
we obtain 
\begin{equation}
k(r_0)=0. 
\end{equation}

\subsection{Initial surface}
\label{Sec:IC}

We summarise the conditions to impose at the initial time surface below.
Hereafter, we describe initial values by using subscript ``0". 
We assume that the LTB region is infinitesimal at the initial time, that is,
\begin{equation}
a(t_{-0})f(\chi_0)=R(t_{+0},r_0). 
\end{equation}
In section~\ref{eqs},
we assumed that the dust fluid passes through the shell without any friction,
and this requirement implied at Eq.~(\ref{equalrho})
that the energy density of the dust fluid is continuous at the shell location, that is,
\begin{equation}
\rho_{-0}=\rho_{+0}=:\rho_0.
\end{equation}
As for the expansion rate, in Eq.~(\ref{contH}) we showed that the LTB region and the outermost EdS region share the common value:
\begin{equation}
H_{+0}=H_{\rm EdS}(t_{+0})=:H_0. 
\end{equation}
On the other hand, $\rho_{+0}$ is not necessarily equal to $\rho_{\rm EdS0}$, and $H_{-0}$ is not necessarily equal to $H_0$. 
To describe discrepancies of them, we introduce $\delta_\rho$ and $\delta_H$ as
\begin{eqnarray}
\delta_\rho&:=&\frac{\rho_0-\rho_{\rm EdS0}}{\rho_{\rm EdS0}}, \\
\delta_H&:=&\frac{H_{-0}-H_0}{H_0}. 
\end{eqnarray}

In the bubble nucleation scenario, the energy density inside the shell is slightly lower than that outside the shell as long as the tension of the shell is positive, as we can see from Eq.~(\ref{ceq4}). 
Then, if we take a time slice on which the Hubble parameter is uniform, 
the Hubble equation~(\ref{rhoeq_interior}) requires the spatial curvature $K$ must be negative.
Based on this observation, we
assume that the inside region is an open dust-dominated FLRW universe
with $K=-1$ and 
\begin{eqnarray}
R(t_{+0},r_0)&=&R_0=br_0, \\
t_{\rm B}(r_0)&=&0, \\
m(r_0)&=&m_0,
\end{eqnarray}
where 
$b$ is a constant that fixes the residual gauge degree of freedom in  $r$, 
which was not fixed by Eq.~\eqref{rgauge}.
Solutions do not depend on the value of $b$, while we need to choose it
so that $r>0$ is maintained throughout the region of interest since
numerical calculation breaks down at $r=0$ (see e.g.\ Eq.~\eqref{eq:dmda}). 
$t_{\rm B}(0)$ has been set to 0 using the time-shift degree of freedom. 
From the Einstein equation \eqref{Eeq}, we obtain 
\begin{eqnarray}
H_0^2&=&\left.\frac{(\del_t R)^2}{R^2}\right|_{t=t_{+0}}
=\left.\frac{2M}{R^3}\right|_{t=t_{+0}}
=\frac{1}{3b^3}m_0
\quad\Leftrightarrow \quad
m_0=3b^3H_0^2. 
\end{eqnarray}

Initial values of the variables can be expressed in terms of $\delta_\rho$ and $\delta_H$ as follows.
From the Hubble equation \eqref{rhoeq_interior}, 
we obtain 
\begin{equation}
\frac{1}{a_0^2}=H_{-0}^2-\frac{8\pi}{3}\rho_{0}
\quad\Leftrightarrow\quad
a_{0}=\frac{1}{H_0\sqrt{\delta_H^2+2\delta_H-\delta_\rho}}, 
\end{equation}
where we have used $H_0^2=8\pi \rho_{\rm EdS0}/3$. 
From the first junction condition \eqref{ceq1}, 
we obtain 
\begin{equation}
\sinh \chi_0 =\frac{R_0}{a_0}=R_0H_0\sqrt{\delta_H^2+2\delta_H-\delta_\rho}. 
\end{equation}
Integrating the Hubble equation \eqref{rhoeq_interior} with respect to $a$, we find the initial time inside the shell $t_{-0}$ to be 
\begin{equation}
t_{-0}=\frac{1}{H_0}\int^1_0
\frac{\sqrt{u}\dd u}{\sqrt{1+\delta_\rho+u(\delta_H^2+2\delta_H-\delta_\rho)}},
\end{equation}
while the integral of Eq.~(\ref{rhoeq_exterior}) with respect to $R$ fixes that for the shell exterior region $t_{+0}$ as
\begin{equation}
t_{+0}=\frac{2}{3H_0}. 
\end{equation}

The values of $y_0$, $\delta_\rho$, $\delta_H$ and $\sigma$ must be fixed so that 
constraint equations \eqref{exdtR-2} and \eqref{exdrR-2} are satisfied. 
At the initial time, these equations can be written as
\begin{eqnarray}
H_0-H_{0-}&=&4\pi \sigma a_0 y_0 , 
\label{eqdelh}
\\
\cosh \chi_0-1&=&4\pi R_0\sigma \sqrt{1+a_0^2y_0^2}. 
\label{eqy}
\end{eqnarray}
Solving these equations for $\sigma$ and $y_0$, we obtain
\begin{eqnarray}
\sigma^2&=&\frac{R_0^2H_0^2(2\delta_H-\delta_\rho)+2
-2\sqrt{1+R_0^2H_0^2(\delta_H^2+2\delta_H-\delta_\rho)}}
{16\pi^2R_0^2}, \\
y_0&=&-\frac{H_0\delta_H}{4\pi\sigma a_0}
\cr
&=&\frac{-R_0H_0^2\delta_H\sqrt{\delta_H^2+2\delta_H-\delta_\rho}}
{\sqrt{R_0^2H_0^2(2\delta_H-\delta_\rho)+2
-2\sqrt{1+R_0^2H_0^2(\delta_H^2+2\delta_H-\delta_\rho)}}}.
\label{y0indelta}
\end{eqnarray}
For a real solution of $\sigma$ to exist, 
the following inequality must be satisfied:
\begin{equation}
R_0H_0>\frac{2\delta_H}{2\delta_H-\delta_\rho}. 
\end{equation}
We can also solve Eq.~\eqref{eqdelh} for 
$\delta_\rho$ as follows:
\begin{equation}
\delta_\rho=2\delta_H-\frac{16\pi^2\sigma^2}{H_0^2}
-\frac{2}{R_0H_0^2}\sqrt{16\pi^2\sigma^2+\delta_H^2H_0^2}.  
\end{equation}
Only three parameters are independent 
among $R_0H_0$, $\sigma/H_0$, $\delta_H$ and $\delta_\rho$. 
In the next section, we specify $R_0H_0$, $\sigma/H_0$ and one of $\delta_H$ or $\delta_\rho$ as the input parameters, and determine the other one in terms of the others using the equations above.

\section{Results}
\label{results}

A bubble nucleation naturally realizes the 
interior region whose energy density is slightly lower compared to the exterior region. 
Below,
we focus on the cases in which 
the Hubble parameter is uniform on the 
initial time slice,
that is,
$\delta_H=0$. 
This condition implies $y=w=\dot m=0$ 
at the 
initial time 
as we can see from Eqs.~\eqref{eq:dotm} and (\ref{y0indelta}), 
and thus $r$ derivatives of physical quantities, e.g., $\del_r m=\dot m/w$,
cannot be numerically calculated there although it is a regular quantity. 
Therefore, in our calculation we set very small but finite value $\delta_H=10^{-8}$.
We confirmed that the numerical results are insensitive to its precise value as long as it is sufficiently small.
We examine also the $\delta_\rho=0$ case in Appendix~\ref{App:rhoconst}.

Before showing the results, we define $H_{\rm hc}$ as the 
Hubble parameter at the horizon crossing
when the shell area radius 
$R(t_+^{\rm s}, r^{\rm s})$ satisfies
\begin{equation}
1/R(t_+^{\rm s}, r^{\rm s})=H_{\rm EdS}(t_+^{\rm s})=:H_{\rm hc}. 
\label{horizoncrossing}
\end{equation}
The Hubble radius is given by $1/H_{\rm hc}$. 
We regarded $R_0H_0$ and $\sigma/H_0$ as the independent parameters 
in the initial data construction,
but it turns out that the analysis is simplified if we regard the value of 
$\sigma/H_{\rm hc}$ as one of the controlling parameters.
Therefore, for each value of $R_0H_0$,
we have tuned the value of $\sigma/H_0$ so that the desired value of 
$\sigma/H_{\rm hc}$ is realized. 
Values of $\sigma/H_0$ and labels for each model 
used in the numerical analysis 
are summarised in Table~\ref{parameter}. 
\begin{table}[htbp]
\caption{Parameter sets for numerical calculations 
in the $\delta_H=10^{-8}$ cases.
The value in each cell shows $\sigma/H_0$ for which $\sigma/H_{\rm hc}$ takes the value specified in the top row when $R_0H_0$ is fixed to the value shown in the leftmost column. 
}
\label{parameter}
\begin{tabular}{|c||c|c|c|}
\hline
\backslashbox{$R_0H_0$}{$\sigma/H_{\rm hc}$}
&$1.72\times 10^{-5}$&$8.58\times 10^{-4}$&$8.58\times 10^{-3}$\\
\hline\hline
2&$4.000\times 10^{-6}$ [{\tt R2-1}]&$2.000\times 10^{-4}$ [{\tt R2-2}]&$2.000\times 10^{-3}$ [{\tt R2-3}]\\
\hline
3&$1.510\times 10^{-6}$ [{\tt R3-1}]&$7.505\times 10^{-5}$ [{\tt R3-2}]&$7.450\times 10^{-4}$ [{\tt R3-3}]\\
\hline
4&$7.000\times 10^{-7}$ [{\tt R4-1}]&$3.490\times 10^{-5}$ [{\tt R4-2}]&$3.445\times 10^{-4}$ [{\tt R4-3}]\\
\hline
\end{tabular}
\end{table}

\begin{figure}[!htbp]
\begin{center}
\includegraphics[scale=0.9]{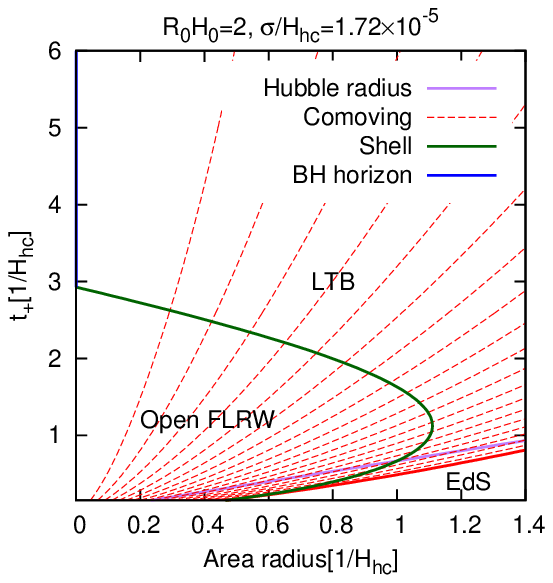}
\includegraphics[scale=0.9]{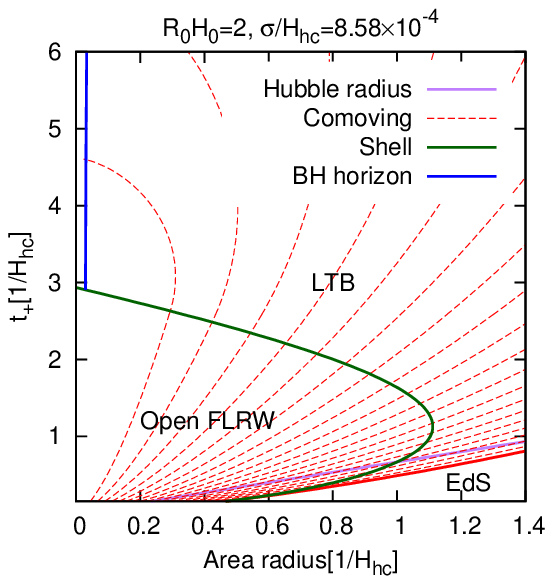}
\includegraphics[scale=0.9]{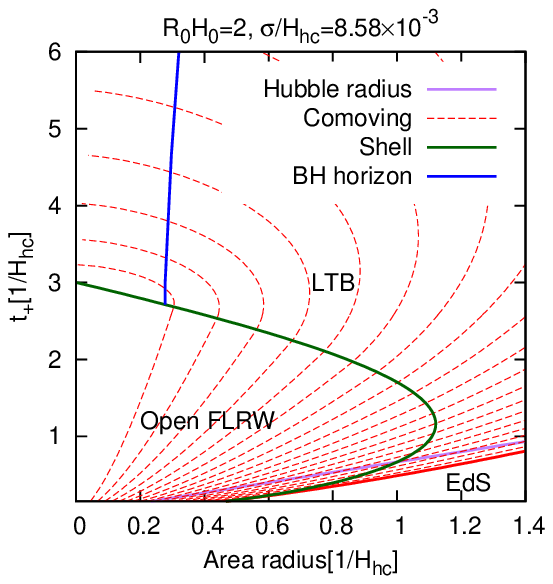}
\includegraphics[scale=0.9]{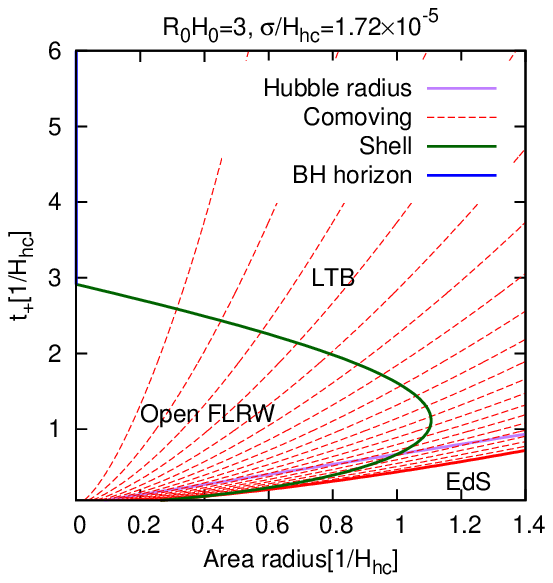}
\includegraphics[scale=0.9]{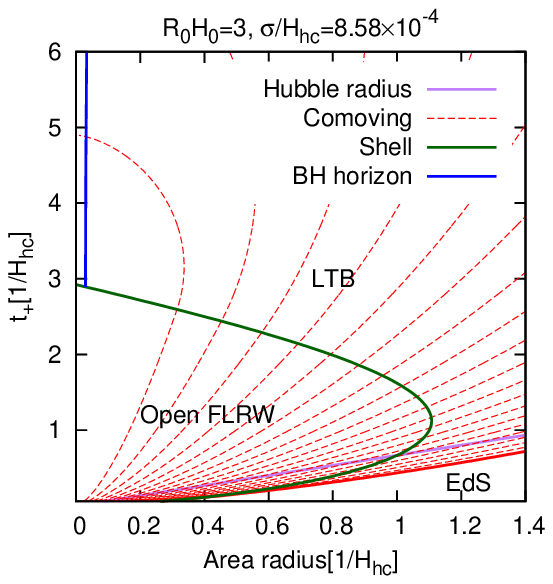}
\includegraphics[scale=0.9]{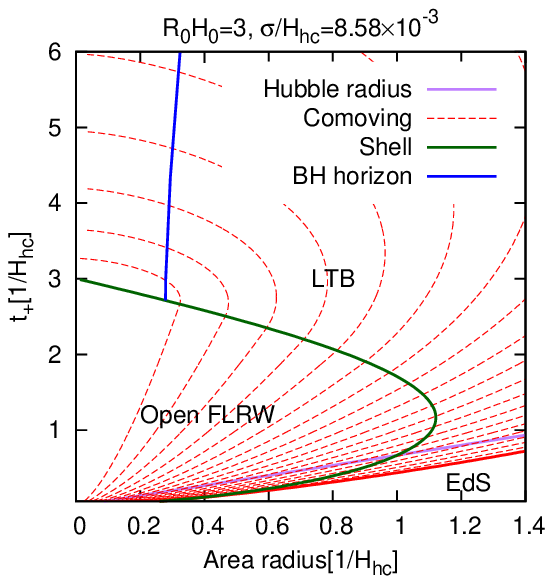}
\includegraphics[scale=0.9]{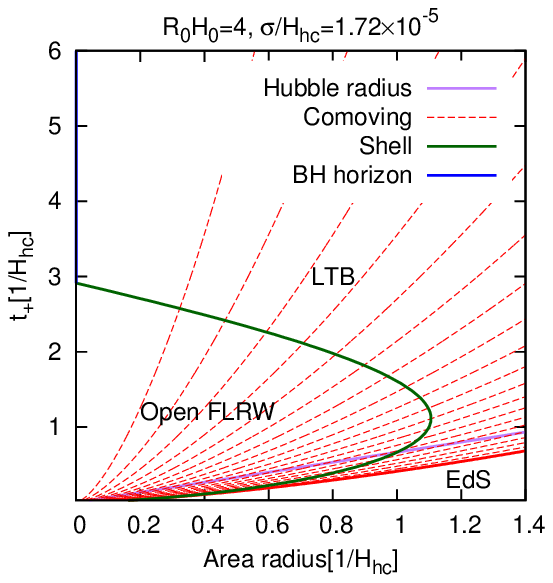}
\includegraphics[scale=0.9]{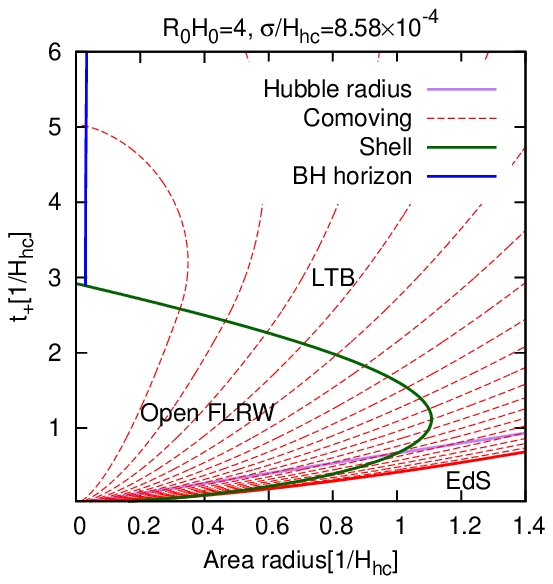}
\includegraphics[scale=0.9]{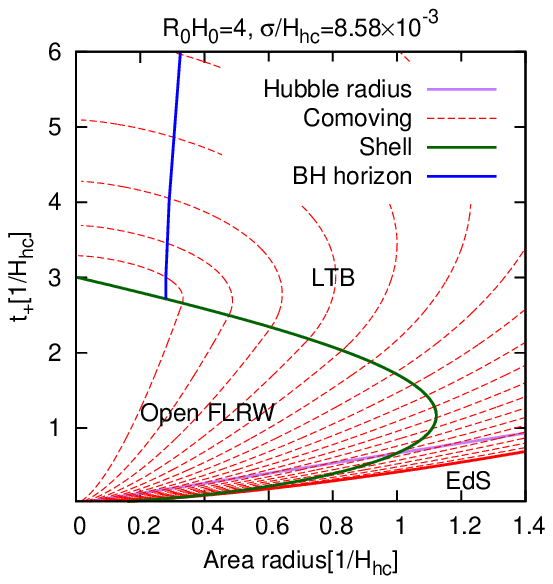}
\caption{
Time evolution of the shell trajectory (green solid curve) and comoving lines of the dust fluid (red dashed curves).
The outermost comoving line separates the LTB region outside the shell 
and the outermost EdS region.
After crossing with the shell trajectory, the dust trajectories are inflected inward and eventually swallowed by the black hole (purple solid line) in the central region of the collapse.
Black hole horizons almost overlap with the vertical axes
in the figures in the left column.
}
\label{fig:spacetime}
\end{center}
\end{figure}

To clarify the possible parameter region of $\sigma/H_{\rm hc}$, 
let us consider the shell at the moment of maximum area radius $\dot R=0$, 
for which the mass of the shell is roughly estimated as $\sim 4\pi\sigma R^2$.
As will be shown below, the maximum area radius of the shell is approximately given by $\sim 1/H_{\rm hc}$ (see e.g.\ Fig.~\ref{fig:spacetime}), hence the mass of the shell at this moment is $\sim 4\pi\sigma/H_{\rm hc}^2$.
Since the shell energy cannot be larger than the total energy inside 
the bubble at the initial time, we obtain the following inequality:
\begin{equation}
\frac{4\pi \sigma}{H_{\rm hc}^2} < M_{\rm max}:=
\frac{4}{3}\pi R_0^3 \rho_{\rm EdS0}=\frac{1}{2}H_0^2R_0^3. 
\label{Mmax}
\end{equation}
If we roughly approximate the shell trajectory before the moment of $\dot R=0$ 
by a comoving trajectory in the EdS universe, 
we obtain 
$$
H^2(t_+^{\rm s})R^3(t_+^{\rm s}, r^{\rm s})\sim {\rm const.}\sim 
H_0^2R_0^3,
$$
where the left-hand side becomes $\sim  1/H_{\rm hc}$ if evaluated at the moment of horizon crossing defined by Eq.~(\ref{horizoncrossing}).
Therefore, from Eq.~\eqref{Mmax}, we find 
\begin{equation}
\frac{\sigma}{H_\text{hc}} \lesssim \frac{1}{8\pi}. 
\label{smax1}
\end{equation}

\begin{figure}[htbp]
\begin{center}
\includegraphics[scale=1.2]{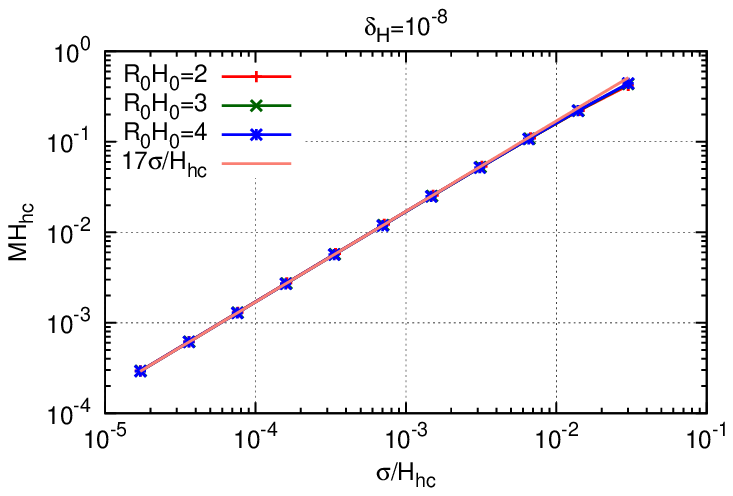}
\caption{
Dependence of the initial black hole mass $M_{\rm BH}|_\text{initial}$ on the shell tension $\sigma/H_{\rm hc}$ for $R_0 H_0 = 2,3,4$.
Fitting the numerical data, the black hole mass is found to behave $M_{\rm BH}|_\text{initial}\simeq 17\sigma / H_\text{hc}^2$ for any $R_0H_0$.
}
\label{msig}
\end{center}
\end{figure}

We summarise the time dependence of the shell and dust fluid motion in 
Fig.~\ref{fig:spacetime}. 
In every case, a black hole forms as a result of the shell collapse. 
where the black hole (apparent) horizon is specified by 
the condition $R(t_+^{\rm s}, r^{\rm s}) =2M(r^{\rm s})$ (see, e.g., Ref.~\cite{Plebanski:2006sd}). 
We can see also that the cases with common value of $\sigma/H_{\rm hc}$ 
share almost the same time dependence.
This feature can be observed also in the value of black hole mass $M_{\rm BH}$
at the black hole formation time as shown in Fig.~\ref{msig}. 
Fitting the numerical data, we find that 
the black hole mass at the formation time 
for all values of $R_0H_0$ is roughly given by 
\begin{equation}
\left.M_{\rm BH}\right|_{\rm initial} \simeq 17\sigma/H^2_{\rm hc}. 
\end{equation}

After the shell collapse, the dust fluid in the LTB region falls into the black hole as shown in Fig.~\ref{fig:spacetime}, and the black hole mass increases with time as a result.
Since only the dust in the LTB region can fall into the black hole,
the maximum value of the black hole mass is 
equal to the initial total mass of the shell and the dust in that region
given by $M_{\rm max}$. 

\begin{figure}[htbp]
\begin{center}
\includegraphics[scale=1.2]{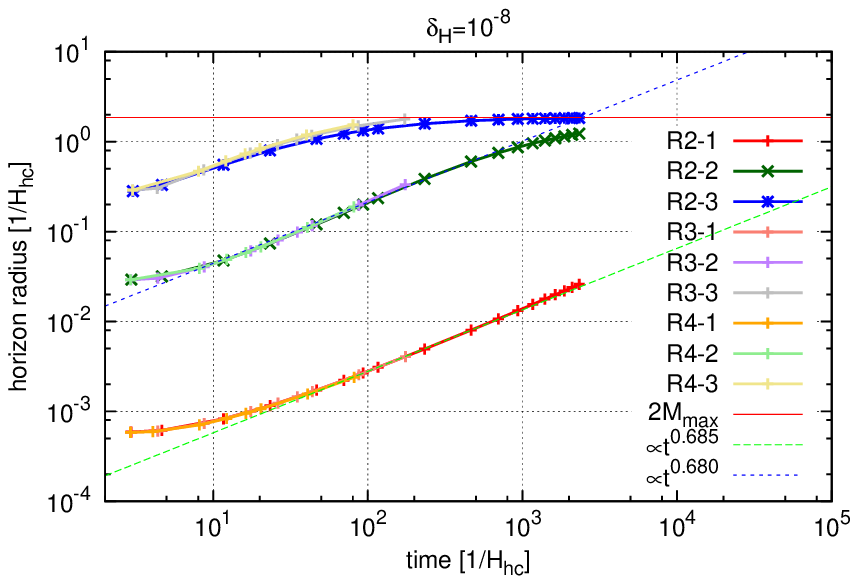}
\caption{Time evolution of the black hole horizon radius.
The maximum value of the black hole radius is shown by the red horizontal line, and the horizon radius converges into this line at the late time.
Fitting the numerical data to a function $\propto t_+^p$ over $t_+  H_\text{hc}\in[50,500]$, we find $p=0.685$ and $0.680$ for the cases {\tt R2-1} and {\tt R2-2}, respectively. The data for different value of $R_0H_0$ ({\tt R3-1}, {\tt R3-2}, {\tt R4-1}, {\tt R4-2}) appear to be consistent with this fitting formula.
}
\label{fig:massevouh}
\end{center}
\end{figure}

The time evolution of the black hole mass is shown in Fig.~\ref{fig:massevouh}. 
When $\sigma/H_\text{hc}\ll 1$, the time evolution is controlled only by the parameter $\sigma/H_{\rm hc}$. 
This plot also indicates that the black hole mass grows obeying a power law in the time domain shortly after the shell collapse before the mass approaches the maximum value (\ref{Mmax}).
Fitting the numerical data of the horizon radius to a function $\propto t^p$ over $t_+ H_\text{hc}\in[50,500]$, we find $p=0.685$ and $0.680$ for the cases {\tt R2-1} and {\tt R2-2}, respectively. 
We can find good agreement also between these power-law functions and 
the data for different values of $R_0H_0$ ({\tt R3-1}, {\tt R3-2}, {\tt R4-1}, {\tt R4-2}).
It would be interesting to confirm if this behavior is a universal one, 
and if it is the case 
we can estimate the time scale for the dust in the LTB region to be 
completely swallowed by the black hole based on this behavior.

\begin{figure}[!htbp]
\begin{center}
\includegraphics[scale=1.]{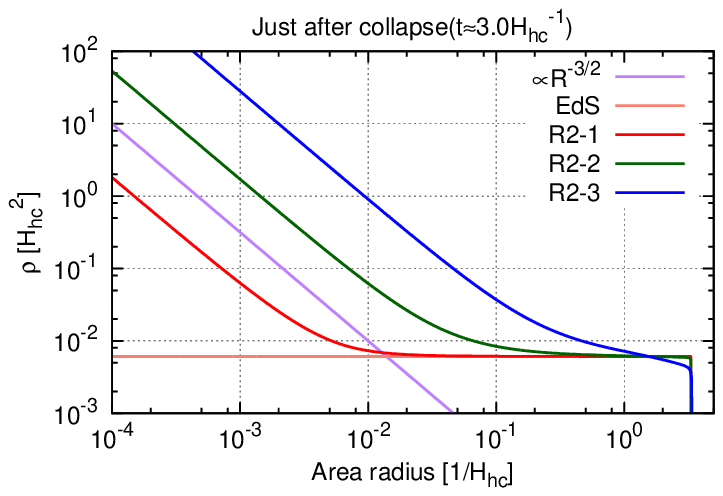}
\includegraphics[scale=1.]{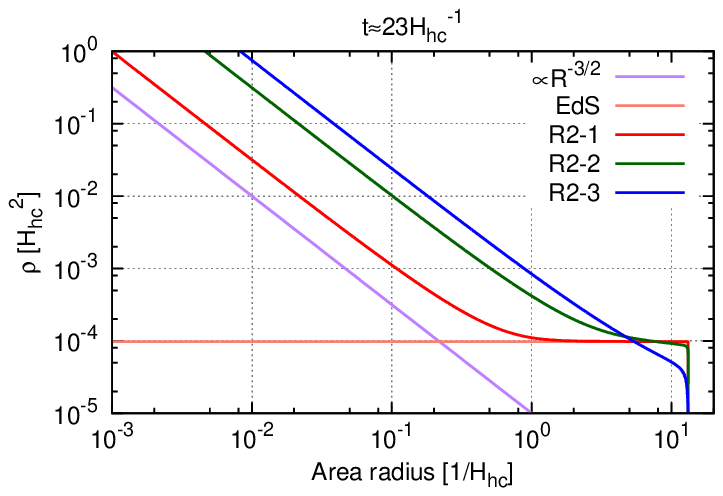}
\includegraphics[scale=1.]{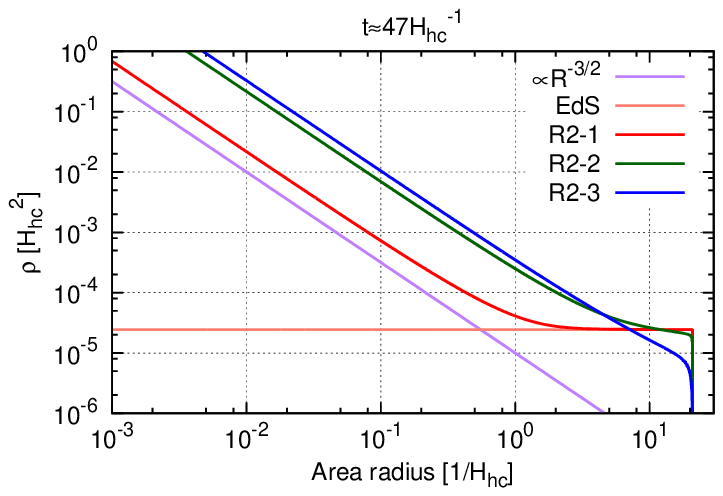}
\includegraphics[scale=1.]{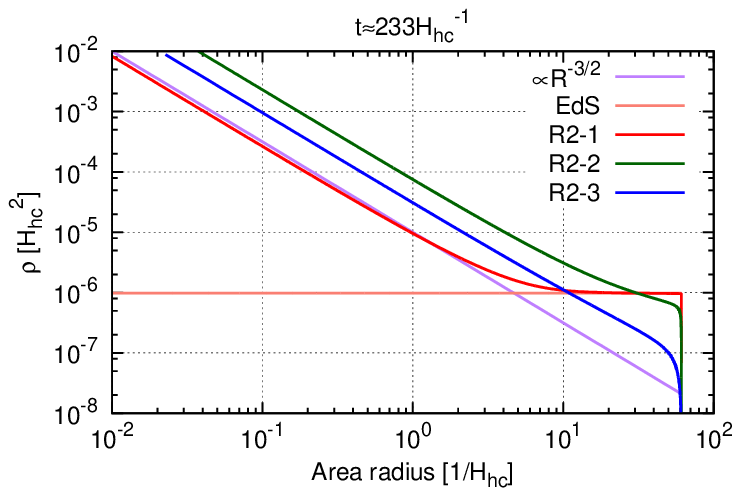}
\caption{Density of the dust fluid in the LTB region 
at $t_+ H_+ = 3.0$, 23, 47, 233
for the cases {\tt R2-1}, {\tt R2-2}, {\tt R2-3}. The cases with different values of $R_0H_0$ give mostly the same results.
At any $t_+$, the density behaves as $\rho\propto R^{-3/2}$ in the region where the area radius is sufficiently small.
}
\label{fig:rho_uh}
\end{center}
\end{figure}

Let us focus on the density profile outside the black hole. 
Since the results do not significantly depend on the value of $R_0H_0$, 
we show our results for only $R_0H_0=2$ cases. 
In Fig.~\ref{fig:rho_uh}, 
we show the density profile in the LTB region at 
$t_+ H_\text{hc}=3.0$, 23, 47, 233 for $R_0H_0=2$. 
From this result we find 
\begin{equation}
\rho\propto R^{-3/2} ~~{\rm near~the~center}.   
\end{equation}
This density profile is expected to be realized 
around a black hole. 
It would provide a significant observable 
relic of the spherical domain wall collapse.

\section{Summary and Discussion}
\label{summary}

In this paper, we have analysed spherical domain wall collapse 
in a dust-dominated universe. 
An under-dense region is surrounded by a spherical domain wall,
and the spacetime region outside the domain wall is described by the LTB solution which is made inhomogeneous due to gravitational interaction with the collapsing domain wall.
This system is supposed to be a relic of a bubble nucleation during the inflation. 

Applying Israel's junction conditions 
to this system, we derived ordinary differential equations 
for the shell trajectory and the inhomogeneity in the LTB region. 
Numerically solving these equations, we found that 
a black hole forms as a consequence of the shell collapse 
for any initial condition.
We also found that the mass of the black hole 
at its formation is
given by $M_{\rm BH}\simeq 17\sigma /H^2_{\rm hc}$, 
where $\sigma$ and $H_{\rm hc}$ are 
the shell tension and the Hubble parameter at the moment of the 
horizon crossing of the shell. 
The matter density $\rho$ in the LTB region near the central region shows 
a characteristic behavior $\rho\propto R^{-3/2}$,
where $R$ is the area radius. 

From the formula $M_{\rm BH}\simeq 17\sigma /H^2_{\rm hc}$,
the black hole mass at its formation
is estimated as 
\begin{equation}
M_{\rm BH}\sim
3.4
\times 10^{17}\left(\frac{\sigma\hbar^2}{{\rm GeV}^3}\right)
\left(\frac{1/H_{\rm hc}}{30\,{\rm kpc}}\right)^{2} M_\odot,  
\end{equation}
where the range of $\sigma/H_{\rm hc}$ is restricted in the region
$\sigma/H_{\rm hc}\lesssim {\cal O}(1)$ due to 
the energy conservation(see Eq.~\eqref{smax1}). 
If we consider 100 Mpc comoving length of a bubble 
at the horizon entry, 
$(\sigma \hbar^2)^{1/3}$ cannot be lager than $\sim 1$ GeV. 
Furthermore, a shell with $(\sigma \hbar^2)^{1/3} \sim 1\,\text{MeV}$ and 100 Mpc 
comoving scale forms a black hole with the mass $\sim 10^9M_\odot$. 
$(\sigma \hbar^2)^{1/3}$ must be larger than this value for the inflation model 
that gives the bubble nucleation to be compatible with the BBN scenario, 
but such a shell would form a black hole heavier than $\sim 10^9M_\odot$, 
which may not be acceptable observationally.
Therefore our setting is not 
likely to be applicable to the late matter-dominated era. 
If there were any matter-dominated epoch 
in the early universe before the matter-radiation equality, 
however, the bubble scale at the horizon crossing could be smaller than the value above.
Black holes with realistic mass could be formed due to the bubble collapse in such a case.

It would be interesting to analyse a bubble collapse in the radiation-dominated epoch, since bubbles of smaller scale enters in the horizon in this epoch.
Such small-scale bubbles could form black holes with realistic mass as long as they collapse to sufficiently small size.
To examine such a scenario, we would need to solve the dynamics of spherically-symmetric spacetime and matter fields numerically. 
This analysis is an important future work
that would also provide further observational features of the bubble nucleation in the early universe.

\section*{Acknowledgements}
We are grateful to H.\ Ishihara, N.\ Kaloper,
M.\ Kimura, K.\ Nakao and M.\ Sasaki for helpful discussions and comments. 
N.T.\ was supported in part by the European Research Council grant no.\ ERC-2011-StG 279363-HiDGR.

\appendix

\section{Figures for $\delta_\rho=0$}
\label{App:rhoconst}

We summarise the results for the $\delta_\rho=0$ cases in this appendix.
The parameter sets and their labels  are listed in 
Table~\ref{parameter_ud}.
The results for this case turn out to be parallel to those in the $\delta_H=10^{-8}$ case.

\begin{figure}[!htbp]
\begin{center}
\includegraphics[scale=0.9]{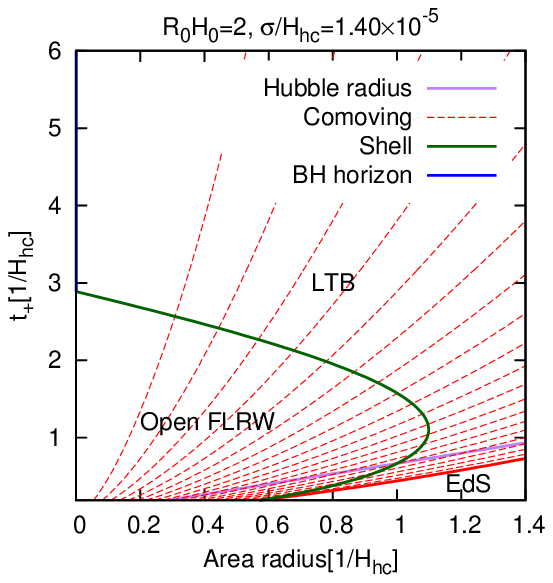}
\includegraphics[scale=0.9]{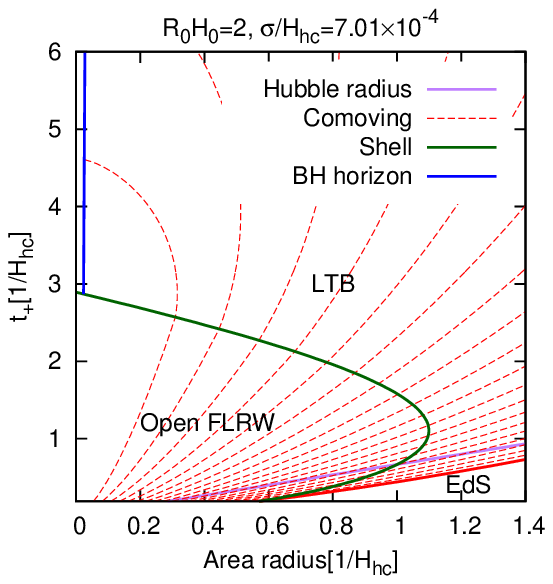}
\includegraphics[scale=0.9]{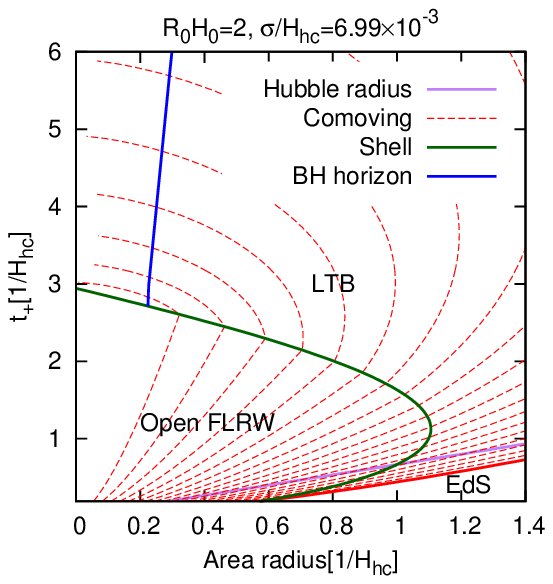}
\includegraphics[scale=0.9]{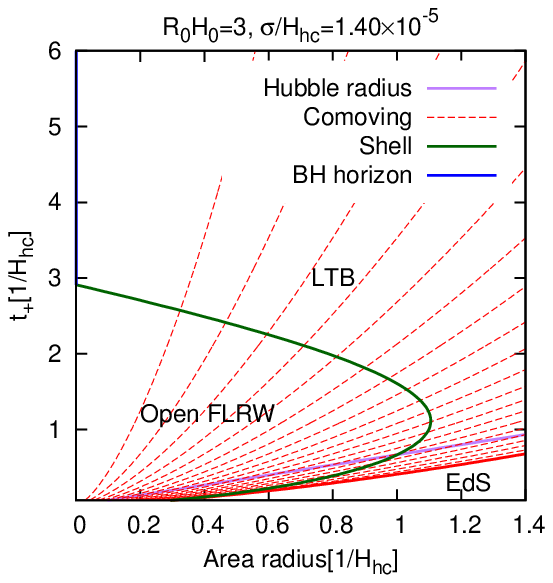}
\includegraphics[scale=0.9]{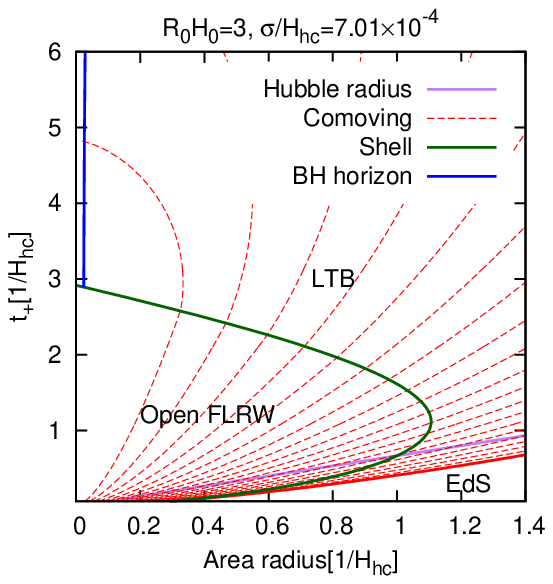}
\includegraphics[scale=0.9]{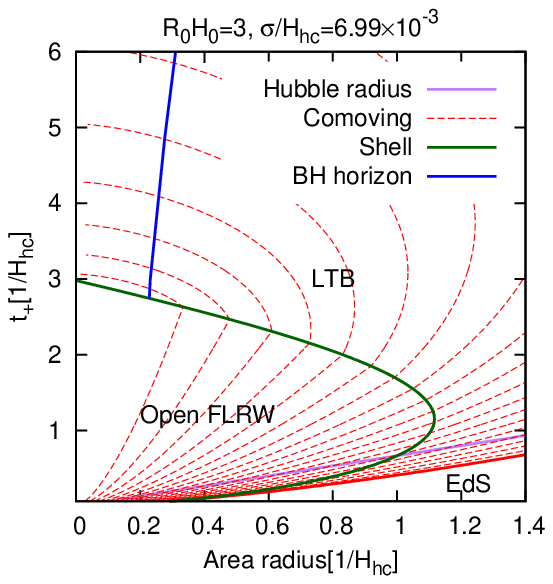}
\includegraphics[scale=0.9]{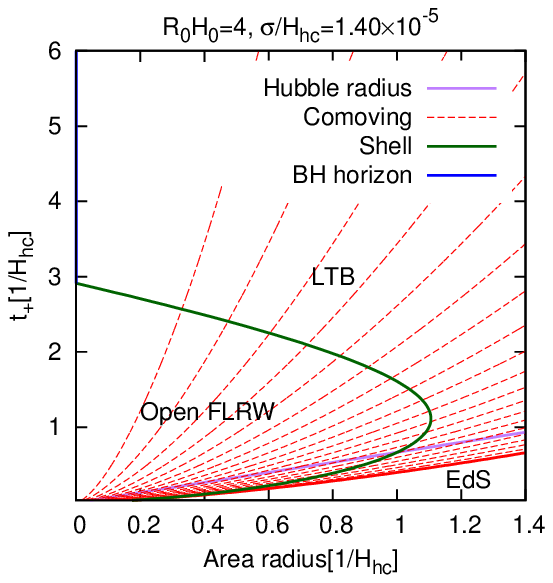}
\includegraphics[scale=0.9]{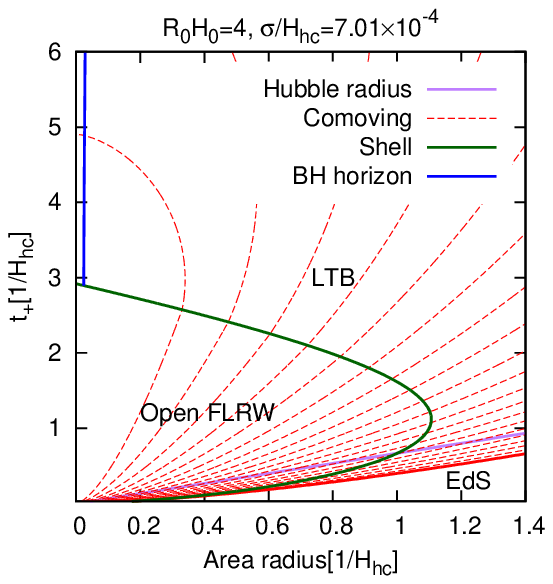}
\includegraphics[scale=0.9]{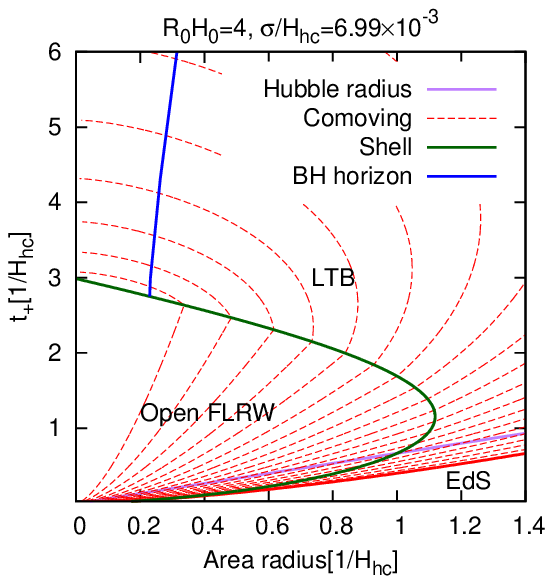}
\caption{
Time evolution of the shell trajectory (green solid curve) and comoving lines of the dust fluid (red dashed curves) in the $\delta_\rho=0$ case.
}
\label{fig:spacetime_ud}
\end{center}
\end{figure}

Figure~\ref{fig:spacetime_ud} shows the trajectories of the shell and dust fluid, and their qualitative behaviors are similar to those in the $\delta_H=10^{-8}$ case shown in Fig.~\ref{fig:spacetime}.

\begin{figure}[htbp]
\begin{center}
\includegraphics[scale=1.2]{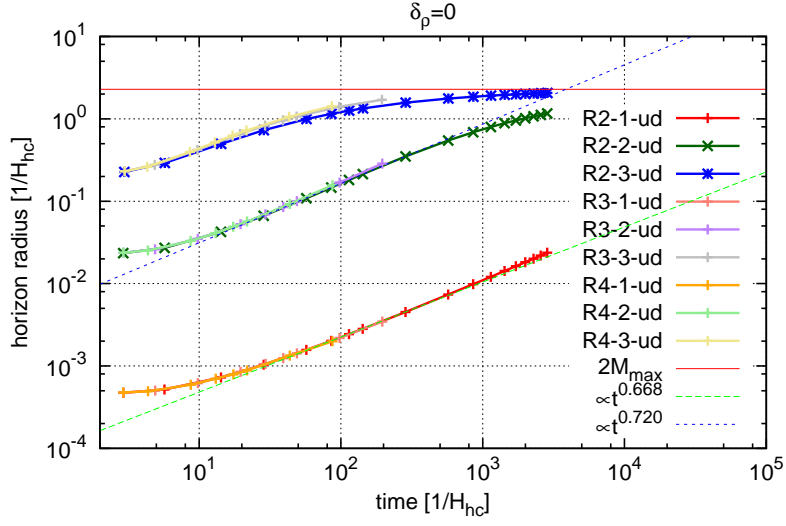}
\caption{Time evolution of the black hole mass in the $\delta_\rho = 0$ case.
The data for common value of $\sigma/H_\text{hc}$ show almost the same time evolution irrespective of the values of $R_0H_0$.
Fitting of a function $\propto t_+^p$ to the numerical data 
over $t_+ H_\text{hc}\in[50,500]$
gives $p=0.668$ and $0.720$ for the cases {\tt R2-1-ud} and {\tt R2-2-ud}, respectively. 
This fitting formula appears to be consistent also with the other results ({\tt R3-1-ud}, {\tt R3-2-ud}, {\tt R4-1-ud}, {\tt R4-2-ud}).
}
\label{fig:massevoud}
\end{center}
\end{figure}

Figure~\ref{fig:massevoud} shows the time evolution of the black hole mass after the collapse. Fitting the numerical data to a function $\propto t_+^p$ over $t_+ H_\text{hc}\in[50,500]$, we find $p=0.668$ and $0.720$ for the cases {\tt R2-1-ud} and {\tt R2-2-ud}, respectively. This fitting formula appears to be consistent also with the other results ({\tt R3-1-ud}, {\tt R3-2-ud}, {\tt R4-1-ud}, {\tt R4-2-ud}).
These powers coincide with those in the $\delta_H = 10^{-8}$ cases within 10\% relative differences.

\begin{figure}[htbp]
\begin{center}
\includegraphics[scale=1.2]{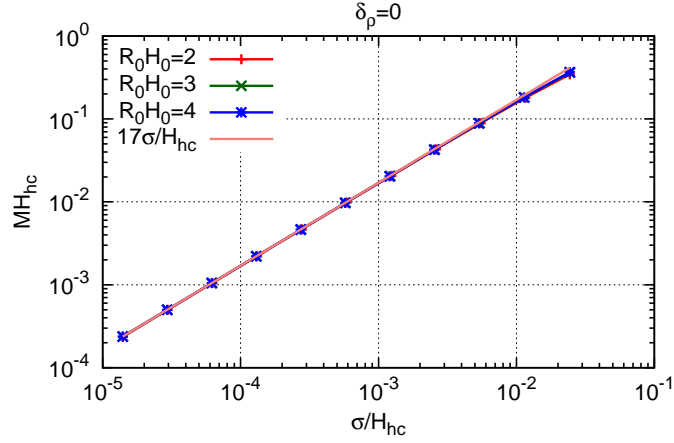}
\caption{
$\sigma/H_{\rm hc}$ dependence of 
the black hole mass $M_{\rm BH}$ for $R_0H_0=2$, 3, 4 in the $\delta_\rho=0$ case.
Irrespective of $R_0 H_0$, the black hole mass obeys $M_{\rm BH} \simeq 17\sigma/H_\text{hc}$.
}
\label{msig_ud}
\end{center}
\end{figure}

\begin{figure}[!htbp]
\begin{center}
\includegraphics[scale=1.]{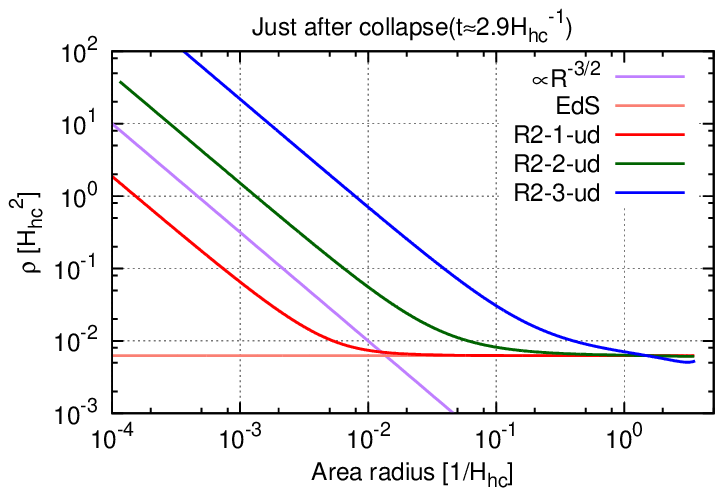}
\includegraphics[scale=1.]{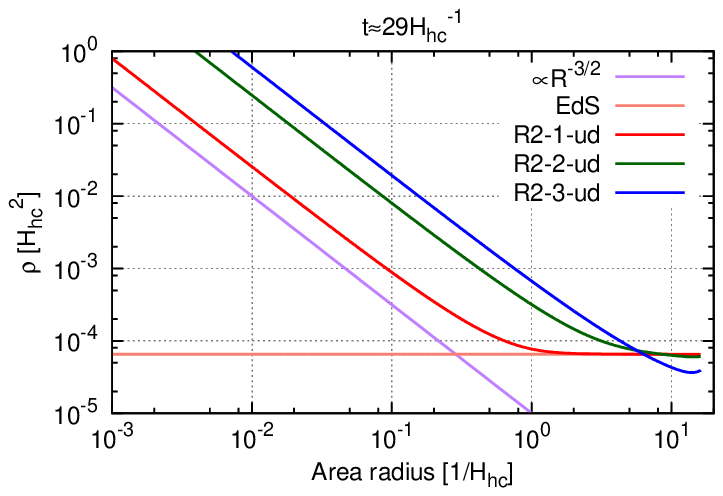}
\includegraphics[scale=1.]{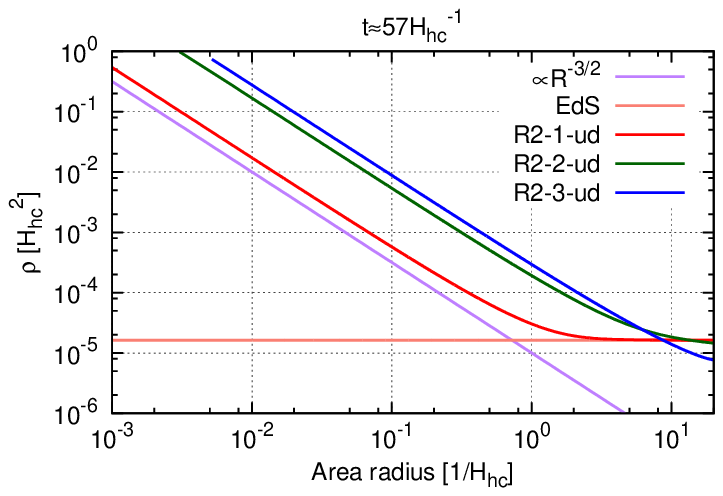}
\includegraphics[scale=1.]{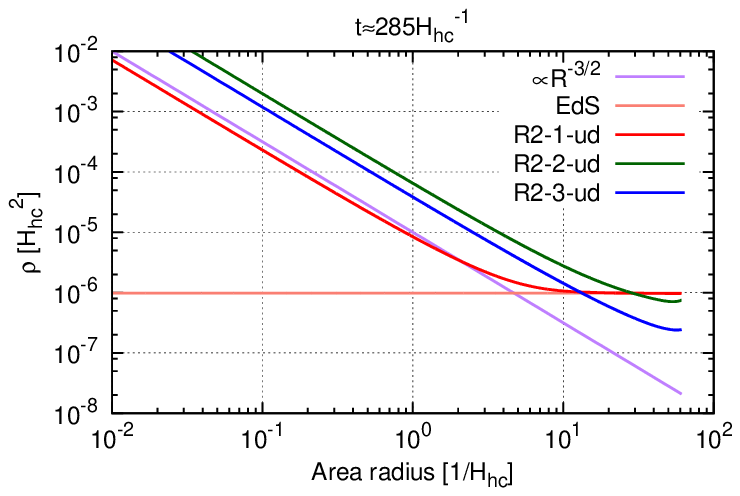}
\caption{
Dust fluid density in the LTB region 
at $t_+ H_\text{hc}=2.9$, 29, 57, 285
in the $\delta_\rho=0$ case.
The density behaves as $\rho\propto R^{-3/2}$ in the region where the area radius $R/H_c$ is sufficiently small.
}
\label{fig:rho_ud}
\end{center}
\end{figure}

Figure~\ref{msig_ud} shows the $\sigma/H_\text{hc}$ dependence of the black hole mass at the moment of its formation. Similarly to the $\delta_H=10^{-8}$ case, the results are independent of $R_0 H_0$ and show the universal behavior $M_\text{BH}\simeq 17 \sigma / H_\text{hc}$.
Figure~\ref{fig:rho_ud} shows the dust density profile in the LTB region at some moments after the shell collapse. The results do not depend on $R_0H_0$ as mentioned above, and thus the results only for $R_0 H_0=2$ are shown there. 
We can see that the universal behavior $\rho\propto R^{-3/2}$ for the $\delta_H=10^{-8}$ cases holds also for these $\delta_\rho=0$ cases.

\begin{table}[htbp]
\caption{Parameter sets for the $\delta_\rho=0$ case.
The value in each cell shows $\sigma/H_0$ for which $\sigma/H_{\rm hc}$ takes the value shown in the top row when $R_0H_0$ is fixed to the value in the left column.
}
\label{parameter_ud}
\begin{tabular}{|c||c|c|c|}
\hline
\backslashbox{$R_0H_0$}{$\sigma/H_{\rm hc}$}
&$1.40\times 10^{-5}$&$7.01\times 10^{-4}$&$6.99\times 10^{-3}$\\
\hline\hline
2&$4.000\times 10^{-6}$ [{\tt R2-1-ud}]&$2.000\times 10^{-4}$ [{\tt R2-2-ud}]&$2.000\times 10^{-3}$ [{\tt R2-3-ud}]\\
\hline
3&$1.370\times 10^{-6}$ [{\tt R3-1-ud}]&$6.870\times 10^{-5}$ [{\tt R3-2-ud}]&$6.810\times 10^{-4}$ [{\tt R3-3-ud}]\\
\hline
4&$6.100\times 10^{-7}$ [{\tt R4-1-ud}]&$3.055\times 10^{-5}$ [{\tt R4-2-ud}]&$3.015\times 10^{-4}$ [{\tt R4-3-ud}]\\
\hline
\end{tabular}
\end{table}

\section{Consistency of the constraint equations}
\label{App:constraints}

Some of the equations derived from the junction conditions turn out to be constraint equations, and they must be kept satisfied on the entire shell trajectory. Below, we show that the constraint equations~(\ref{ceq2}) and (\ref{ceq3}) are kept satisfied by the other equations and the initial data given in the main text.%
\footnote{
This derivation is equivalent to showing that two among Eqs.~(\ref{ceq3}), (\ref{deq4}), (\ref{deq5}) and (\ref{deq6}) are not independent from the others if the first junction condition~(\ref{ceq1}) is assumed to hold on the whole trajectory of the shell. This degeneracy is a consequence of the Bianchi identity since Eqs.~(\ref{ceq3}), (\ref{deq4}), (\ref{deq5}) and (\ref{deq6}) can be regarded as two ``dynamical'' equations, the ``Hamiltonian'' and ``momentum'' constraints for the ``time evolution'' in the direction of the unit normal $s^\mu$, respectively.
}

The total differential of \eqref{ceq2} gives a 
dynamical equation: 
\begin{eqnarray}
&&\frac{\dd}{\dd \tau}\left[\gamma_- \del_t (\ln a)  +y \del_\chi (\ln f) 
-\gamma_+\del_t (\ln R)-w \del_r (\ln R)\right]=0
\\
&\Leftrightarrow&
\left(\frac{\del_\chi f}{f}+\frac{a\del_t a y}{\gamma_-}\right)\dot y
-\left(\frac{\del_r R}{R}+\frac{\del_t R w}{R\gamma_+}\right)\dot w
+\frac{wr^2}{2R\del_r R}\dot k\cr
&&+\left(-\frac{(\del_t a)^2}{a^2}+\frac{\del_t^2 a}{a}\right)\gamma_-^2
+\left((\del_t a)^2-\frac{1}{f^2}\right)y^2
+\left(\frac{(\del_t R)^2}{R^2}-\frac{\del_t^2 R}{R}\right)\gamma_+^2\cr
&&+\left(\frac{2\del_r R\del_t R}{R^2}-\frac{\del_t\del_r R}{R}\right)\gamma_+w
+\left(\frac{(\del_r R)^2}{R^2}+\frac{kr}{R\del_r R}\right)w^2=0, 
\label{deq2-0}
\end{eqnarray}
where we have used $\del_\chi^2 f=-Kf$, $(\del_\chi f)^2=1-Kf^2$ 
and 
\begin{eqnarray}
\dot \gamma_-&=&a\del_t a y^2+a^2y\dot y /\gamma_-, \\
\dot \gamma_+&=&w\dot w/\gamma_+.  
\end{eqnarray}
We also used the fact that 
the differential of $\del_r R$ with respect to $\tau$ is
calculated as 
\begin{equation}
\frac{\dd }{\dd \tau}(\del_r R)
=\frac{\dd }{\dd \tau}\sqrt{1-kr^2}
=\frac{ -2kr w-r^2\dot k}{2\del_r R}. 
\end{equation}
As shown in Ref.~\cite{Yoo:2010qn}, 
$\del_t\del_r R$ can be expressed as 
\begin{eqnarray}
\del_t\del_r R&=&G\del_r m+H\del_r k +I \del_r t_{\rm B}+J\cr
&=&\frac{1}{w}(G\dot m+H\dot k+I\dot t_{\rm B})+J.
\end{eqnarray}
Eliminating $\del_t\del_r R$ using this equation,
Eq.~\eqref{deq2-0} is reduced to
\begin{eqnarray}
&&\left(\frac{\del_\chi f}{f}+\frac{a\del_t a y}{\gamma_-}\right)\dot y
-\left(\frac{\del_r R}{R}+\frac{\del_t R w}{R\gamma_+}\right)\dot w
+\left(\frac{wr^2}{2R\del_r R}-\frac{\gamma_+ H}{R}\right)\dot k
-\frac{\gamma_+ G}{R}\dot m-\frac{\gamma_+ I}{R}\dot \tb \cr
&&
+\left(-\frac{(\del_t a)^2}{a^2}+\frac{\del_t^2 a}{a}\right)\gamma_-^2
+\left((\del_t a)^2-\frac{1}{f^2}\right)y^2
+\left(\frac{(\del_t R)^2}{R^2}-\frac{\del_t^2 R}{R}\right)\gamma_+^2\cr
&&+\left(\frac{2\del_r R\del_t R}{R^2}-\frac{J}{R}\right)\gamma_+w
+\left(\frac{(\del_r R)^2}{R^2}+\frac{kr}{R\del_r R}\right)w^2=0. 
\label{deq2}
\end{eqnarray}
The total differential 
of Eq.~(\ref{ceq3})
gives another dynamical equation:
\begin{eqnarray}
&&\frac{\dd}{\dd \tau}\left[w\del_t (\ln R)+\gamma_+\del_r (\ln R)
-y\del_t a-\frac{\gamma_-}{a}\del_\chi (\ln f)\right]
=0,\\
&&\Leftrightarrow
-\left(\del_t a+\frac{a \del_\chi f y }{f\gamma_- }\right)\dot y
+\left(\frac{\del_t R}{R}+\frac{\del_r R w}{R\gamma_+}\right)\dot w
\cr
&&
+\left(-\frac{\gamma_+r^2}{2R\del_r R}+\frac{w H}{R}\right)\dot k
+\frac{w G}{R}\dot m+\frac{w I}{R}\dot \tb 
\cr
&&
+\frac{\del_t a \del_\chi f}{a^2f}\gamma_-^2
-\frac{\del_t a \del_\chi f}{f}y^2
+\left(-\del_t^2a+\frac{1}{af^2}\right)\gamma_-y
\cr
&&
-\frac{\del_t R \del_r R}{R^2}\gamma_+^2
+\left(\frac{ J}{R}-\frac{\del_r R \del_t R}{R^2}\right)w^2
\cr
&&
+\left(-\frac{(\del_r R)^2+(\del_t R)^2}{R^2}+\frac{\del_t^2 R}{R}-\frac{kr}{R\del_r R}\right)\gamma_+w
=0.  
\label{deq3}
\end{eqnarray}

Using the expressions for $\dot m$, $\dot k$, $\dot \tb$ and $\dot y$ derived from Eqs.~\eqref{deq1}, \eqref{deq4}, \eqref{deq5}, \eqref{deq6} and the constraints shown in the main text, we can show that the time derivative of the constraint equations~(\ref{deq2-0}) and (\ref{deq2}) are trivially satisfied.
In the derivation, we have used the following identities:
\begin{eqnarray}
&&CI-EG=\frac{r^3}{6R}, \label{eq:use1}
\\
&&DI-EH=-\frac{1}{2}r^2, \label{eq:use2}
\\
&&E=-\del_t R, \label{eq:use3}
\\
&&I=\frac{r^3m}{6R^2}, \label{eq:use4}
\\
&&FI+EJ-kr=\frac{r^3m\del_r R}{6R^2}-\frac{r^2m}{2R}. \label{eq:use5}
\end{eqnarray}

In section~\ref{Sec:IC}, we assumed that Eqs.~(\ref{exdtR-2}) and (\ref{exdrR-2}), which are derived from the constraints (\ref{exdtR}) and (\ref{exdrR}) and thus from Eqs.~(\ref{ceq2}) and (\ref{ceq3}), to hold at the initial time. Thus we can conclude that the constraints (\ref{ceq2}) and (\ref{ceq3}) are satisfied at the initial time and are guaranteed to be kept satisfied due to the derivations above.


\end{document}